\documentclass[aps,prb,twocolumn,superscriptaddress,floatfix,longbibliography]{revtex4}
\usepackage{amsfonts}
\usepackage{amssymb}
\usepackage{graphicx}
\usepackage{dcolumn}
\usepackage{bm}
\usepackage{amsmath}
\usepackage[colorlinks,linkcolor=magenta,citecolor=blue,urlcolor=blue]{hyperref}
\usepackage{changes}

\begin{document}

\title{Localization transition, spectrum structure and winding numbers for one-dimensional non-Hermitian quasicrystals}

\author{Yanxia Liu}
\affiliation{Beijing National Laboratory for Condensed Matter Physics, Institute of
Physics, Chinese Academy of Sciences, Beijing 100190, China}
\author{Qi Zhou}
\email{qizhou@nankai.edu.cn}
\affiliation{Chern Institute of Mathematics and LPMC, Nankai University, Tianjin 300071,
China}
\author{Shu Chen}
\email{schen@iphy.ac.cn}
\affiliation{Beijing National Laboratory for Condensed Matter Physics, Institute of
Physics, Chinese Academy of Sciences, Beijing 100190, China}
\affiliation{School of Physical Sciences, University of Chinese Academy of Sciences,
Beijing, 100049, China}
\affiliation{Yangtze River Delta Physics Research Center, Liyang, Jiangsu 213300, China}
\date{\today }

\begin{abstract}

By analyzing the Lyapunov exponent (LE), we develop a rigorous, fundamental
scheme for the study of  general non-Hermitian quasicrystals with both complex
phase factor and non-reciprocal hopping.  Specially, the localization-delocalization
transition point, $\mathcal{PT}$-symmetry-breaking point and the winding number transition points are
determined by LEs of its dual Hermitian model. The analysis was based
on Avila's global theory, and we found that  winding
number is directly related to the acceleration, the slope of the LE, while quantization of acceleration is the
crucial ingredient of Avila's global theory.
This result applies as well to the models with higher winding, not only the simplest Aubry-Andr\'{e} model.
As typical examples, we obtain the analytical phase boundaries of localization transition for non-Hermitian
Aubry-Andr\'{e} model in the whole parameter space, and the complete phase diagram is straightforwardly
determined. For the non-Hermitian Soukoulis-Economou model, a high winding model,
we show how the phase boundaries of localization transition and winding number transitions
relate to the LEs of its dual Hermitian model.
Moreover, we discover an intriguing feature of robust spectrum, i.e.,
the spectrum keeps invariant when one changes the complex phase parameter
$h$ or non-reciprocal parameter $g$ in the region of $h<|h_c|$ or $g<|g_c|$ if the
system is in the extended or localized state, respectively.

\end{abstract}

\maketitle




\section{Introduction}

Localization induced by disorder is an
old but everlasting research topic in condensed matter physics \cite{anderson1958absence}.
While Anderson localization induced by random
disorder are thoroughly studied \cite{abrahams1979scaling,lee1985disordered,evers2008anderson,Thouless72}, localization
transition in quasiperiodic systems has attracted increasing interest in
recent years \cite%
{luschen2018,Aubry1980,Kohmoto1983,Thouless1988,roati2008}. In comparison with the
random disorder systems, the quasiperiodic systems manifest some peculiar properties and may support exact results
due to the existence of duality relation for the transformation between
lattice and momentum spaces. A typical example is the Aubry-Andr\'{e}
(AA) model \cite{Aubry1980}, which undergoes a
localization transition when the quasiperiodical potential strength
exceeds a transition point determined by a self-duality condition \cite{Jitomirskaya1999}.
Various extensions of AA models have been studied \cite{Kohmoto1983,Ceccatto,Zhou2013,Cai,DeGottardi,Kohmoto2008,WangYC-review,Chandran,Chong2015}.
The quasiperiodic lattice models can support energy-dependent mobility edges when either short-range (long-range) hopping processes \cite{biddle2011localization,biddle2010predicted,ganeshan2015nearest,li2016quantum,li2017mobility,li2018mobility,DengX}
or modified quasiperiodic potentials \cite{sarma1988mobility,sarma1990localization,YCWang2020} are introduced.

The interplay of non-Hermiticity and disorder brings new perspective for the localization phenomena.
Due to the releasing of the Hermiticity constrain, non-Hermitian random matrices contain much
 more rich symmetry classes according to Bernard-LeClair classification
\cite{BL,HYZhou,CHLiu,Sato} than the corresponding Hermitian
Altland-Zirnbauer classification. In the scheme of random matrix theory, it has
been demonstrated that the spectral statistics for
non-Hermitian disorder systems displays many different features from the
Hermitian systems \cite{Goldsheid,Markum,Molinari,Chalker,Shindou,Ueda-RM,XRWang,Ryu}.
The interplay of the nonreciprocal hopping and random disorder has
been studied in terms of the Hatano-Nelson-type
models \cite{hatano1996localization,hatano1998non,kolesnikov2000localization,Gong,ZhangDW,Hughes}.
The effect of complex disorder potentials has also been investigated \cite{HuangYi,ZhangDW2,Tzortzakakis}.
Non-Hermitian quasiperiodic systems have also attracted intensive studies very recently \cite{Yuce,longhi2019metal,jazaeri2001localization,jiang2019interplay,zeng2019topological,longhiPRL,
ZengQB,Zeng2020,Liutong,Liu2020,Liu2020L,Cai2021AA,Longhi2021AA,Zhai2020,Liu2020DM,
Tang2021TO,Cai2021PW}.

The LE is an important quantity to characterize the localization properties of disorder systems and plays an essential role in the study of localization transition. As one of Avila's Fields Medal work, he developed  global theory of quasiperiodic cocycles and studied the  delicate but fundamental property of LE. This is quite an important progress in the spectral theory of self-adjoint quasiperiodic Schr\"{o}dinger operators \cite{Avila2015,Avila2017}. Nevertheless, the application of the global theory to the study of physical properties of quasiperiodic systems is not well recognized in the physics communities. Particularly, the study of non-Hermitian quasicrystals in terms of the global theory was only addressed very recently \cite{Liu2020L}, and a systematic scheme applicable for general non-Hermitian quasiperiodic systems is not established yet.
The studies of the general non-Hermitian quasiperiodic models with high winding number are neglected
due to the lack of the exact transition points and universal formulas. In this paper, we develop a fundamental scheme for the study of general non-Hermitian quasiperiodic systems by applying  Avila's global theory, where the non-Hermitian systems
can be realized by introducing both non-reciprocal
hopping and complex phase factor in the Hermitian quasiperiodic model. We find some universal
results to determine the localization-delocalization transition point, $\mathcal{PT}$-symmetry-breaking point and the winding number transition points.
For the case in presence of complex phase factor, the picture of LE actually gives us the mechanism of how winding number and
localized phase change with complex phase factor.  It is surprising that the relevant information for the non-Hermitian systems can be acquired from
their  dual Hermitian  models. For the case in presence of non-reciprocal hopping, the skin effect-localization  transition  and winding number transition
are also directly related to LEs.

We stress that our theory and formalism are valid for general non-Hermitian quasiperiodic systems.
For a better understanding, our general theory is made concrete by focusing on two typical examples, i.e., non-Hermitian
AA model and Soukoulis-Economou model, as showcases for presenting the main results. Despite its deceptively simple form, the phase
boundaries of localization-delocalization transition of the general non-Hermitian
AA model are still not known, except of two limit cases in the absence of either
non-reciprocal hopping \cite{longhiPRL} and complex potential \cite{jiang2019interplay}.
A complete phase diagram with analytical phase boundaries
in the full parameter space is lacking.
In addition, although the coincidence of localization transition point
with the $\mathcal{PT}$-symmetry-breaking point in the $\mathcal{PT}$-symmetry AA model
has been numerically observed \cite{longhiPRL}, no analytical proof is given.We shall clarify these issues by applying our general scheme. Some unusual
and unexplored spectrum features of non-Hermitian AA models, i.e., the spectra are invariant
with the change of complex phase parameter $h$ or non-reciprocal parameter $g$ in specific
regions, are also unveiled.  The feature of robust spectra is found to exist very commonly in non-Hermitian quasiperiodic systems.

The paper is organized as follows. In section II, we first introduce the general model and present the formalism of our general theory. We start from the  systems with complex quasiperiodic potential in the subsection II.A, and demonstrate that the localization-delocalization transition point for general non-Hermitian quasicrystals with complex phase factor can be determined by LEs of its dual Hermitian model. Under the general framework, both the complex AA model and Soukoulis-Economou model are studied.
Then we study the general case in the presence of nonreciprocal hopping in the subsection II.B. Taking the non-Hermitian AA model as a typical example, we obtain the
complete phase diagram which is determined by an analytical formula for the localization transition point.
In section III, We study the properties of winding
numbers and relate them directly to the slope of LEs. Then we identify that the phase diagram of non-Hermitian quasiperiodic models can be characterized by winding numbers.
In section IV, we study the properties of robust spectrum and skin effect. The invariance of spectrum structure of non-Hermitian AA model under the change of  $h$ or $g$ in specific regions
are studied and analyzed. Then we study the interplay of skin effect and localization
and demonstrate that the sensitivity of spectrum structures to the change of boundary
condition from periodic to open boundary condition (PBC to OBC) can distinguish
the skin and localized phases. In section V, we give some examples beyond the AA model and Soukoulis-Economou model. A summary is given in the final section.

\begin{figure}[tbp]
\includegraphics[width=0.45\textwidth]{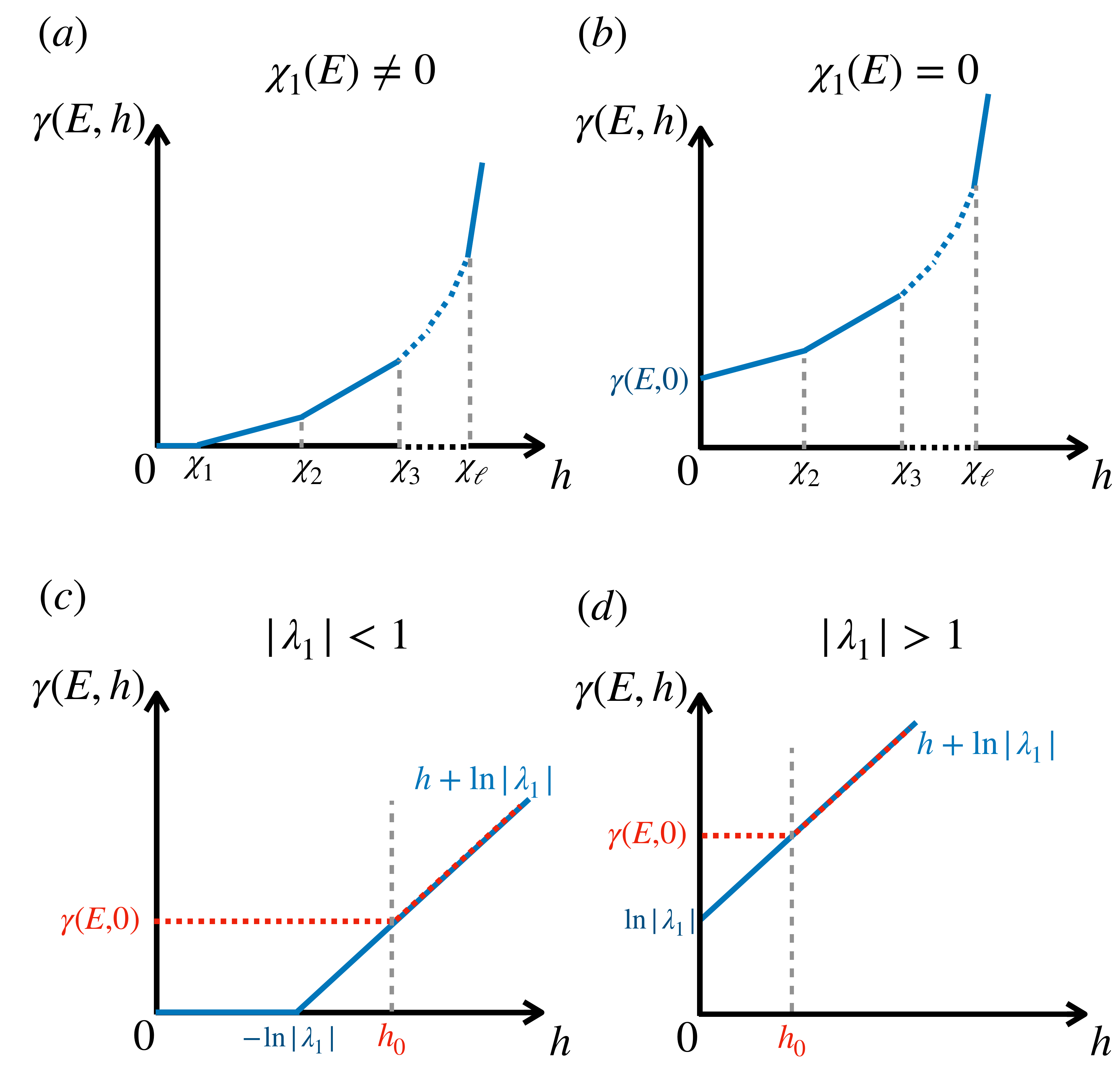}
\caption{ Schematic representation of Eq. (\ref{LEG}): the LE
$\gamma(E,h)$ as a function of $h$ for (a) $\chi\neq 0$ and (b) $\chi=0$, respectively.
For the AA model, the LE $\gamma(E,h)$ with
(c) $|\lambda_1|<1$ and (d) $|\lambda_1|>1$, respectively.}
\label{fig1}
\end{figure}

\section{Models and general theory}
We consider the general non-Hermitian quasiperiodic
models with both complex potential and non-reciprocal hopping,
described by%
\begin{equation} \label{Ham1}
H=\sum_{j=1}^{N}\left( t_{L}|j\rangle \left\langle j+1\right\vert +t_{R}\left\vert
j+1\right\rangle \langle j|+V_{j}|j\rangle \langle j|\right) ,
\end{equation}%
where $t_{L}=te^{-g}$ and $t_{R}=te^{g}$ are the left-hopping and
right-hopping amplitude, respectively, $V_{j}$ is given by
\begin{equation}
V_{j}=\sum_{l=1}^{d} 2 \lambda_l \cos[l(2\pi \omega j+\theta)],
\end{equation}%
with
\[
\theta =\phi +ih
\]
describing a complex phase factor, and $N$ is the lattice size. For convenience,
we set $t=1$ as the unit of energy and take $\omega =\left( \sqrt{5}%
-1\right) /2$, which can be approached by $\omega =\lim_{n\rightarrow \infty
}\frac{F_{n-1}}{F_{n}}$ with the Fibonacci numbers $F_{n}$ defined
recursively by $F_{n+1}=F_{n}+F_{n-1}$ and $F_{0}=F_{1}=1$. By taking $|\psi
\rangle =\sum_{j}u_{j}|j\rangle $, the eigen equation is given by
\begin{equation}
Eu_{j}=e^{-g}u_{j+1}+e^{g}u_{j-1}+V_{j}u_{j},  \label{model}
\end{equation}%
where the eigenvalue $E$ is generally complex.

\subsection{Models with complex quasiperiodic potential}
We first discuss the case in the absence of non-reciprocal hopping, i.e., $%
g=0$. For $\phi=0$, we have $V_j = V^* _{-j}$, the model (\ref{Ham1}) with $%
g=0$ has $\mathcal{PT}$ symmetry\cite{Bender,Yuce}.
Our whole analysis depends on  Avila's global theory of quasi-periodic
Schr\"odinger operator \cite{Avila2015}(see Appendix A for brief introduction), where the key is to
analysis the LE  $\gamma \left( E,h\right)$ with respect to $h$.
The  LE is given by
\begin{equation} \label{LEte}
\gamma \left( E\right) =\lim_{n\rightarrow \infty }\frac{1}{ n} \ln
\left\vert \left\vert T_{n}\left( E \right) \right\vert \right\vert,
\end{equation}%
where the transfer matrix
\begin{equation}
T_{n}\left( E\right) =\prod_{j=1}^{n}T^{j}=\prod_{j=1}^{n}\left(
\begin{array}{cc}
E-V_{j}& -1 \\
1 & 0%
\end{array}%
\right)
\end{equation}%
and  $\left\vert \left\vert A\right\vert \right\vert $ represents the norm of
the matrix $A$, defined by
\[
\left\vert \left\vert A\right\vert \right\vert = max_{i=1:n} \sqrt{\lambda_i (A^T A)}
\]
with $\lambda_i (A^T A)$ being the $i$-th eigenvalue of $A^T A$.

As shown in \cite{Avila2015},
 $\gamma \left( E,h\right)$ is a convex and piece-wise linear function
 with respect to $h$ with their slopes being integers. If $V_j$ is trigonometric polynomial (i.e. $d<\infty$),
 then the extreme points of $\gamma \left( E,h\right)$
 can be determined by the LE of corresponding dual Hermitian Hamiltonian \cite{GJYZ}.
 More precisely, it has the following expansion:
\begin{widetext}\label{lya-formula}
\begin{eqnarray}
\gamma \left( E,h\right)=
\left\{
\begin{array}{cc} \label{LEG}
\gamma(E,0),~~&h \in [0,\chi_{1}(E)],\\
   \vdots  & \vdots \\
\gamma(E,\chi_{i}(E))+ (h-\chi_{i}(E))\sum_{j=1}^{i}{n_{j}},&~~h \in (\chi_i(E),\chi_{i+1}(E)],\\
 \vdots  & \vdots\\
\gamma(E,\chi_{\ell}(E))+(h-\chi_l(E))\sum_{j=1}^{\ell}{n_{j}}, &~~h \in (\chi_\ell(E),\infty),
\end{array} %
\right.
\end{eqnarray}
\end{widetext}
where $0\le \chi_1(E)<\cdots <\chi_\ell(E)$ are the non-negative LEs
with multiplicity $n_1, \cdots, n_\ell$ for
the dual model of the system (\ref{Ham1}) with $h=0$
\begin{equation}\label{zlong}
E\tilde{u}_k=\sum\limits_{l=-d}^{d} \lambda_{|l|}\tilde{u}_{k+l}
+2\cos(2\pi \omega k)\tilde{u}_k.
\end{equation}
This means that the LE  $\gamma \left( E,h\right)$ can be uniquely
determined by $\gamma(E,0)$, $\chi_i$ and $n_i$. These points $h=\chi_i$ are knotted,
which correspond to some physical consequence.
However, we emphasize that duality is not the essence of our approach. The crucial things are the
extreme point and the slope of $\gamma \left( E,h\right)$, and duality only gives an efficient
way to achieve these parameters. That is also the reason why our approach works for general non-Hermitian
quasiperiodic models (c.f. Section V). We also should point out that in the following discussion,
we only need to consider the case $E$ belong to the spectrum of the Hermitian case($h=0$).
Actually, based on Avila's global theory, we can get that if $E$ doesn't belong to the spectrum
of the Hermitian case, $\gamma \left( E,0\right)>0$ and $\chi_1(E)>0$. To better understand
the Eq. (\ref{LEG}),  we show the LE $\gamma \left( E,h\right)$ in Fig \ref{fig1}
(a) and (b) as a function of $h$ with $h>0$.
Note that $\gamma(E,h)$ is
symmetric over $h$, i.e.,
\begin{equation}\label{sym}
\gamma(E,h)=\gamma(E,-h).
\end{equation}

An intriguing issue is that the $\mathcal{PT}$-symmetry breaking transition and
the transition from extended to localized states can be both determined by the LE of
Hermitian dual model $\chi_1(E)$.
To explain it clearly, we will start with  the representative model, the AA model, i.e., only $\lambda_1\neq0$,
which is discussed in detail in the following. The dual model form for the model (\ref{Ham1})
with $g=0$, $h=0$ and $\lambda_{l\geqslant2}=0$ is
\begin{equation} \label{AAdual}
E\tilde{u}_k=\lambda_{1}\tilde{u}_{k+1}+\lambda_{1}\tilde{u}_{k-1}
+2\cos(2\pi \omega k)\tilde{u}_k,
\end{equation}%
The exact LE $\chi \left( E\right)$ of this dual model  can be obtained by Eq. (\ref{LEte})
with the transfer matrix $T_{n}$ is given by
\begin{equation}
T_{n}\left( E\right) =\prod_{j=1}^{n}T^{j}=\prod_{j=1}^{n}\left(
\begin{array}{cc}
\frac{E-2\cos(2\pi \omega k)}{\lambda_1}& -1 \\
1 & 0%
\end{array}%
\right).
\end{equation}%
From the discussions in \cite{Avila2015,YCWang2020}, we have
\begin{equation}
\chi (E)  =  \max \{-\ln |\lambda_1|,0\}, \label{LEchi}
\end{equation}
if the energy $E$ belongs to the spectrum.
Thus the LE for the original model with $h\neq 0$ can be written as
\begin{align}
\gamma \left( E,h\right)=
\left\{
\begin{array}{cc} \label{LEcos}
\gamma \left( E,0\right), &h \in [0,\chi(E)],\\
\gamma \left( E,\chi(E)\right)+(h-\chi(E)), &h \in (\chi(E),\infty),
\end{array} %
\right.
\end{align}
which can be also rewritten as
\begin{equation}
\gamma (E,h)  =  \max \{\ln |\lambda_1 |+|h|,0\}. \label{LE}
\end{equation}
The LE can be also directly derived from the original model (see Appendix A.1)
without the need to introduce the dual model.

Figs.\ref{fig1}(c) and (d)  show  the LE $\protect\gamma(E,h)$
for non-critical AA models with $|\lambda_1|<1$ and $|\lambda_1|>1$, respectively.
Note $\gamma \left( E,h\right) =0$ indicates the extended state, and $\gamma
\left( E,h\right) >0$ corresponds the localized state. Thus if $|\lambda_1|<1$, all the eigenstates of Hermitian AA model ($h=0$) are extended.
When $|h|>0$, there exists a localization-delocalization transition point  determined by
\begin{equation}
|h|=-\ln |\lambda_1|.
\label{ME1}
\end{equation}
If $|\lambda_1|>1$, all the eigenstates of Hermitian
AA model are localized and all the eigenstates stay localized with $|h|>0$.

In the following, we first consider the case $|\lambda_1|<1$.
For simplicity, we only discuss the
case $h>0$. By Avila's global theory \cite{Avila2015},
 $E$ does not lie in the spectrum of the Hamiltonian $h=h_0$, if and only if $\gamma (E,h_0)>0$,
and $\gamma (E,h)$ is a linear function around $h_0$.
Therefore, if $E$ lies in the spectrum of the Hermitian case  $h=0$, it  belongs to the spectrum of the system
with $h<-\ln |\lambda_1|$, but does not belong to the spectrum of the system
with $h>-\ln |\lambda_1|$,  as shown in the blue dashed line of Fig.\ref{fig1} (c).
Conversely, if the  energy $E$ (might be complex) doesn't  lie in the spectrum of
the Hermitian case $h=0$, then $\gamma (E,0)>0$ and
$\gamma (E,h)$ is locally constant in $h$, as shown in the  red dashed line of Fig.\ref{fig1} (c).
Note that $h_0$ is an extreme point of $\gamma (E,h)$ if and only if $h_0> -\ln |\lambda_1|$.
Therefore these energies $E$  do not belong to the spectrum of the system
with $h<-\ln |\lambda_1|$, which might belong to the spectrum of the system
with $h>-\ln |\lambda_1|$.
By the above discussions,  we prove that the extended states
have real energies when $h<-\ln |\lambda_1|$, and
the localized states have complex eigenvalues when $h>-\ln |\lambda_1|$.  This explains why the localization
transition point coincides with the $\mathcal{PT}$-symmetry breaking point. Furthermore, the spectrum keeps
 invariant in the  regime of extended states, which is  indeed a Cantor set by famous result
 of Avila-Jitomirskaya \cite{Avila}.

However, the case $|\lambda_1|>1$ will be much simpler. By similar discussion as above,
one can easily deduce that when $h>0$, all the sates are localized, which have
complex eigenvalues, as shown in Fig.\ref{fig1} (d).

\begin{figure}[tbp]
\includegraphics[width=0.5\textwidth]{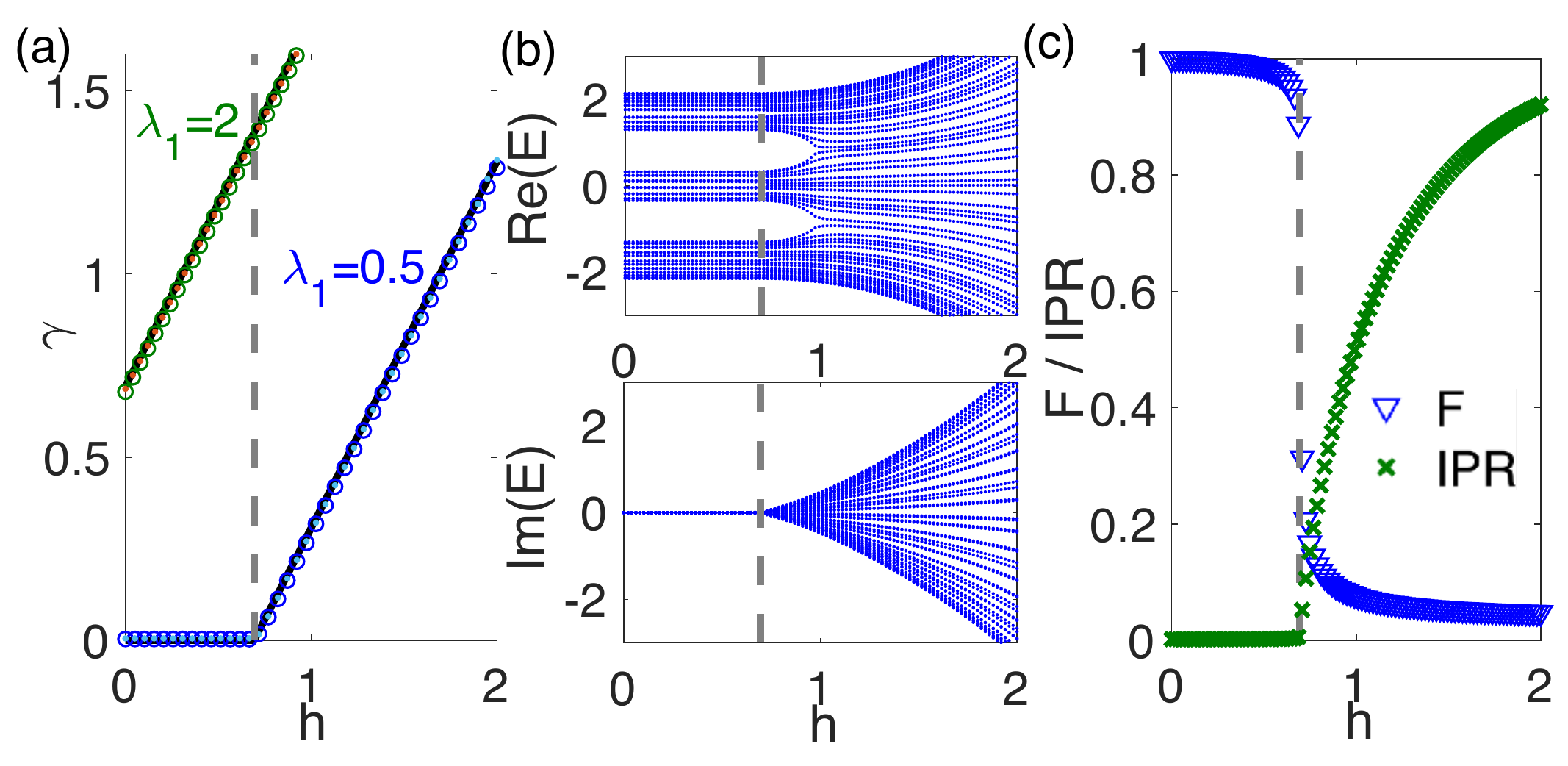}
\caption{(a) Numerical results for the LE of eigenstates corresponding to
the minimum (circles) and maximum(dots) real part of eigenvalues Re$(E)$ versus $h$
for the system with $g=0$, $N=1597$, $\lambda =0.5$ and $2$, respectively.
The black solid lines represent the exact solution of the LE obtained by (\protect\ref{LE}).
(b) The real and the imaginary part of the eigenvalue spectra versus
$h$ for the system with $\lambda_1 =0.5$, $g=0$ and $N=55$.
(c) The fidelity $F$ and IPR of eigenstate corresponding to the minimum real part of eigenvalues Re$(E)$
versus $h$ for the system with $g=0$, $N=1597$, and $\lambda_1 =0.5$.}
\label{fig2}
\end{figure}

To get a straightforward understanding, next we demonstrate numerical results of LE, inverse
participation ratio (IPR), the normalized participation ratio (NPR)\cite{li2016quantum,li2018mobility,li2017mobility} and  energy spectrum as a
function of $h$. For a finite system, the
LE can be numerically calculated via
\begin{equation}
\gamma \left(
E\right) =\ln \left( \max \left( \theta _{i}^{+},\theta _{i}^{-}\right)
\right) , \label{NLE}
\end{equation}
where $\theta _{i}^{\pm }\in \mathbb{R}$ denote eigenvalues of
the matrix
\begin{equation}
\mathbf{\Theta =}\left( T_{N}^{\dag }
T_{N} \right) ^{1/(2N)} . \label{NLE2}
\end{equation}
The IPR and NPR of an eigenstate
is defined as
\[
\text{IPR}^{(n)}=(\sum_{j}\left\vert u_{j}^{n}\right\vert
^{4})/\left( \sum_{j}\left\vert u_{j}^{n}\right\vert ^{2}\right) ^{2},
\]
and
\begin{equation}
{\rm NPR}^{(n)}=\left[N\sum_{i}|u_{i}^{n}|^{4}\right]^{-1}, \label{NPR}
\end{equation}
where the superscript $n$ labels the $n$th eigenstate of system, and $j$
represents the lattice coordinate. For an extended eigenstate, IPR$\simeq 1/N$ approaches zero as
$N\rightarrow \infty $  and NPR is a finite value. On the other hand, IPR$\simeq 1$
and NPR$\simeq 0 $ for a full localized eigenstate.

\begin{figure*}[tbp]
\includegraphics[width=0.9\textwidth]{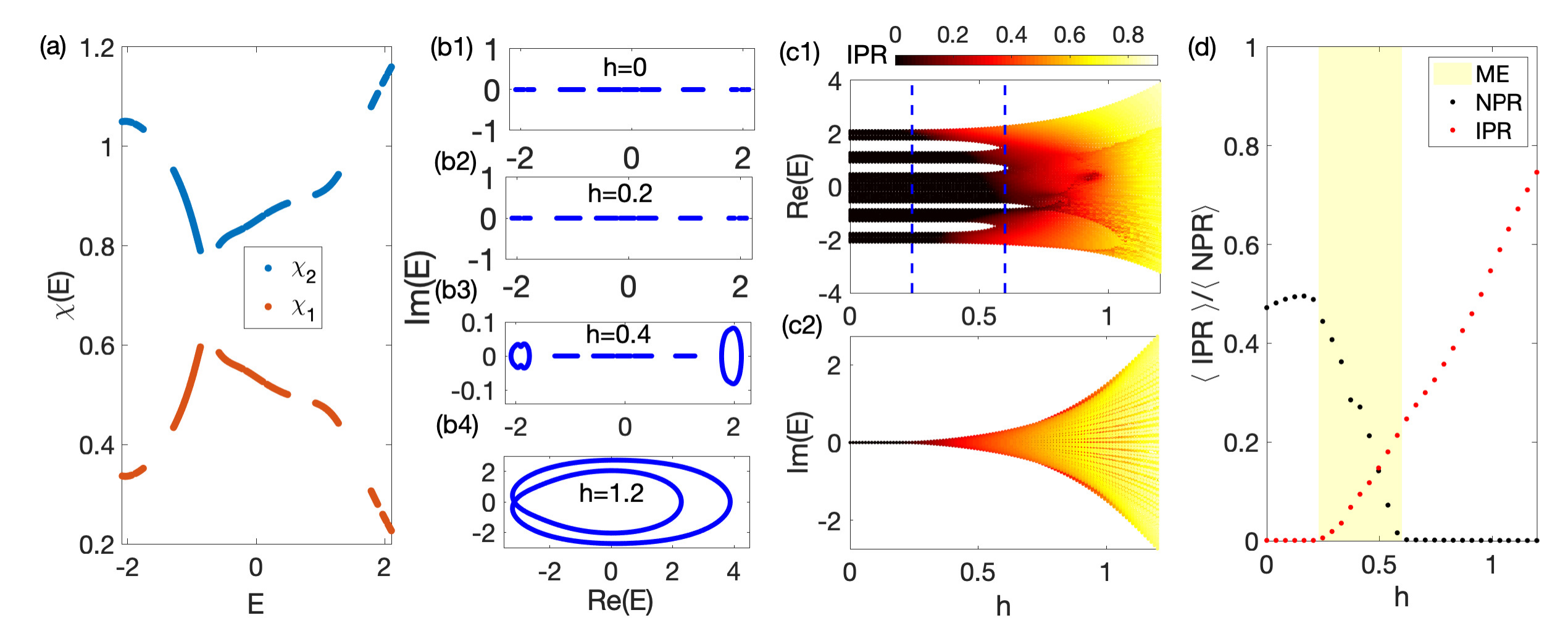}
\caption{(a) The LE of the dual model with $\lambda_1=0.2$ and $\lambda_2=0.25$.
(b) The complex spectrum for systems with (b1) $h=0$, (b2) $0.2$, (b3) $0.4$ and (b4) $1.1$, respectively.
(c)The real and imaginary part of the eigenvalue spectra versus
$h$ for the system with $g=0$, $\lambda_1 =0.2$, and  $\lambda_2=0.25$.
Dashed blue lines indicate the transition point: $h_1=0.23$ and $h_2=0.6$.
(d) Averaged IPR and NPR for all eigenstates in the model.}
\label{fig3}
\end{figure*}

Figs.\ref{fig2} (a) and (c) show the numerical results of the LE and IPR
versus $h$, respectively. If $|\lambda_1|<1$, when $|h|<\chi(E)$, where $\chi(E)=-\ln |\lambda_1|\approx 0.7$ for $\lambda_1
=0.5$, all eigenstates are extended states and both LE and IPR approach zero.
on the other hand,  when $h>\chi(E)$, both LE and IPR have a sudden increase. If $|\lambda_1|>1$, all eigenstates are
localized, as shown in the Fig.\ref{fig2}(a) with $\lambda_1=2$. The numerical results of LE for
finite size system are found to agree well with the analytical result (\ref{LE}).
In Fig.\ref{fig2}(b), we display the real and imaginary part of eigenvalues versus $h$ for the system with $\lambda_1=0.5$.
While all eigenvalues are real for $h<\chi(E)$,  they become complex when $h$ exceeds $\chi(E)$. This clearly shows that the
transition from extended to localized states and $\mathcal{PT}$-symmetry breaking
transition have the same boundary.

It is also interesting to notice that the spectrum does not change with $h$ in the extended region as long as
$h<\chi(E)$. This kind of phenomenon is quite unusual,
since the Hamiltonian with different $h$ is not unitary equivalent, and it's just kind of {\it robust spectrum}.
The similarity of two eigenstates can be characterized by the fidelity
 \begin{equation}
F(h)=\langle \psi_g(0) |\psi_g(h)\rangle,  \label{Fidelity}
\end{equation}
where $\psi_g(h)$ is the  eigenstate of corresponding the minimum real part
of eigenvalue, and $\psi_g(0)$ with $h=0$. The eigenstates vary only slightly from $h=0$ to $h<\chi(E)$,
and change suddenly at  the transition point $h=\chi(E)$,  as shown in Fig. \ref{fig2}(c).

Although, the above phenomenons show in a simple case, these phenomenons can be
commonly found in general non-Hermitian quasiperiodic models \eqref{Ham1} with $g=0$.
If $\chi_1(E)=0$, which means $\gamma(E, 0)>0$, there is no extended-localized
transition for complex phase $h>0$. We thus only need to consider the case $\chi_1(E)>0$.
If $0<h<\chi_1(E)$ for a given eigenvalue $E$ of the system with $h=0$, the eigenvalue remains
unchanged and the corresponding eigenstate is extended.
If $h>\chi_1(E)$, the eigenvalue becomes complex and the corresponding eigenstate is localized.
As a consequence, $h=\min\{\chi_1(E)\}$ gives the beginning of $\mathcal{PT}$ symmetry  breaking and
extended-mixed  transition, while $h=\max\{\chi_1(E)\}$
gives us the mixed-localized transition. That is to say, if $\min\{\chi_1(E)\}<h<\max\{\chi_1(E)\}$,
the mobility edges will occur, the spectrum will have both real and complex energies, and
the phases will be mixture of extended and localized states.
Moreover, if  $ 0< h< \min\{\chi_1(E)\}$,  the spectrum is exactly the same as the Hermitian case $h=0$, thus has
robust spectrum when changing $h$.  Just note that if  $\min\{\chi_1(E)\}=\max\{\chi_1(E)\}$ or say $\chi_1(E)$ is a constant,
then $h=\min\{\chi_1(E)\}$ gives the extended-localized transition point, just as the non-Hermitian AA model.
Finally, we point out $\chi_1(E)$ is also the inverse of localization length of the eigenstate for the Hermitian dual model \eqref{zlong}, and if
$\chi_1(E)>0$, the corresponding eigenstate of system \eqref{zlong} is localized.

Next we will demonstrate this with Soukoulis-Economou model \cite{Soukoulis},
one of the first proposals of one-dimensional quasiperiodic models containing single-particle mobility edges.
It is a tight-binding model with nearest-neighbor
hopping terms as well as two quasiperiodic on-site potentials:
\begin{equation} \label{cos2}
V_j=2\lambda_1 \cos (2\pi \omega j +i h)+ 2 \lambda_2 \cos (4 \pi \omega j + 2i h).
\end{equation}
Our analysis shows that the mobility edge for the system with any $h$ can be
determined by the LE $\chi_1(E)$ for its
dual Hermitian system \eqref{zlong}.

For the potential (\ref{cos2}) with $h=0$, its  dual model \eqref{zlong} is
\begin{equation}\label{Dualcos2}
E\tilde{u}_k= \lambda_{1}(\tilde{u}_{k+1}+\tilde{u}_{k-1})+\lambda_{2}(\tilde{u}_{k+2}+\tilde{u}_{k-2})
+2\cos(2\pi \omega k)\tilde{u}_k,
\end{equation}
and
$\tilde{u}_k=\sum_{n}e^{-i 2 \pi \omega k j}u_j.$
We can rewrite  (\ref{Dualcos2}) as
 \begin{equation*}
\left(
\begin{array}{c}
\tilde{u}_{k+3} \\
\tilde{u}_{k+2}\\
\tilde{u}_{k+1}\\
\tilde{u}_{k}\\
\end{array}%
\right) =T^{k}
\left(
\begin{array}{c}
\tilde{u}_{k+1} \\
\tilde{u}_{k}\\
\tilde{u}_{k-1}\\
\tilde{u}_{k-2}\\
\end{array}%
\right),
\end{equation*}
with
\begin{equation*}
T^{k}=
\left(
\begin{array}{cc}
C_2^{-1}(EI_2-B_{2})&-C_2^{-1}C_2^*\\
I_2 & 0  \\
\end{array}%
\right),
\end{equation*}
where $I_2$ is the $2\times2$ identity matrix,  the matrices $C_2$ and $B_2$ are given by
\begin{equation*}
C_2=
\left(
\begin{array}{cc}
\lambda_2&\lambda_1\\
 0 &\lambda_2\\
\end{array}%
\right),
\end{equation*}
and
\begin{equation*}
B_{2}=
\left(
\begin{array}{cc}
2\cos(2\pi \omega (k+1))& \lambda_1\\
 \lambda_1 &2\cos(2\pi \omega k)\\
\end{array}%
\right).
\end{equation*}
Denote the transfer matrix $T_{n}(E) =\prod_{k}T^{k}$, then the LEs of the model (\ref{Dualcos2}) are given by
\begin{equation} \label{LEtwo}
\chi_i=\ln \theta_i,
\end{equation}
where $\theta_i$ are the eigenvalues of matrix
$$\mathbf{\Theta}=\left( T_{N}^{\dag}
T_{N} \right) ^{1/(2N)}.  $$
Since $T^{k}$  is a $4\times4$ complex symplectic matrix, their LEs  (\ref{LEtwo}) come in pairs, and we write it as
 $ -\chi_2(E)\leq - \chi_1(E) \leq 0\leq \chi_1(E) \leq \chi_2(E)$. The LE (\ref{LEG}) for the original model with $d=2$ and $h\neq 0$ can be written as
\begin{widetext}
\begin{eqnarray}
\gamma \left( E,h\right)=
\left\{
\begin{array}{cc} \label{LEcos2}
\gamma \left( E,0\right), &~~h \in [0,\chi_1(E)],\\
\gamma \left( E,\chi_1(E)\right)+(h-\chi_1(E)), &~~h \in (\chi_1(E),\chi_2(E)],\\
\gamma \left( E,\chi_2(E)\right)+2(h-\chi_2(E)), &~~h \in (\chi_2(E),\infty).
\end{array} %
\right.
\end{eqnarray}
\end{widetext}

Fig. \ref{fig3}(a) shows the LEs $\chi_1(E)$ and $\chi_2(E)$  of the dual model (\ref{Dualcos2}) with $\lambda_1=0.2$
and $\lambda_2=0.25$ as a function of eigennergy $E$, and in Fig.\ref{fig3}(b) and (c), we display its spectrum versus $h$. We also show the averaged IPR and NPR in Fig.\ref{fig3}(d) to distinguish these phases. From these pictures, it is clear that the spectrum keeps invariant in the region of $h<h_1$, where $h_1 =\min\{\chi_1(E)\}=0.23$
is also the $\mathcal{PT}$-symmetry breaking point, all the eigenvalues are
real and all the eigenstates are extended.  The mobility edge region $h_1<h<h_2$  represents the system with a
mixture of localized and extended states, where  $h_2=\max\{(\chi_1(E)\}=0.6$.
The extended states have real eigenvalues, while the localized states have complex eigenvalues,
as shown in Fig.\ref{fig3} (b3) and (c). Eventually, when $h>h_2$, all the eigenstates are localized, and their corresponding eigenvalues are complex.
Our results clearly indicate that the three distinct phases can be divided  by $h=\max\{\chi_1(E)\}$ and $h=\max\{\chi_1(E)\}$.

\subsection{Effect of  non-reciprocal hopping}
Now we consider the general case with $g\neq 0$. The nonreciprocal hopping
breaks the $\mathcal{PT}$ symmetry and may induce skin effect under
OBC. The Hamiltonian $H(g)$ under OBC can be transformed to $H^{\prime }$
via a similar transformation
\begin{equation}
H^{\prime }=SH(g)S^{-1},  \label{smt}
\end{equation}%
where
\[
S=\text{diag} \left( e^{-g},e^{-2g},\cdots ,e^{-Ng}\right)
\]
is a similarity matrix with only diagonal entries and $H^{\prime }=H(g=0)$ is the Hamiltonian with $g=0$. The eigenvectors of $%
H$ and $H^{\prime }$ satisfy $\left\vert \psi \right\rangle
=S^{-1}\left\vert \psi ^{\prime }\right\rangle $. An extended state $%
\left\vert \psi ^{\prime }\right\rangle $ under the transformation $S^{-1}$
becomes skin states, which exponentially accumulate to one of
boundaries \cite{jiang2019interplay,Yao,Kunst,Alvarez,Xiong,Lee}.

A localized state of $H^{\prime }$ generally takes the following form
\[
\left\vert u_{i}\right\vert \propto e^{- \left\vert i-i_{0}\right\vert/\xi
},
\]
where $i_{0}$ represents the position of localization center of a given localized state, $\xi=1/\gamma$ is
the localization length, and $\gamma$ is
the LE of the localized state for the system with $g=0$. Then the corresponding wavefunction of
$H(g)$ takes the following form:
\begin{equation}
\left\vert u_{i}\right\vert \propto \left\{
\begin{array}{cc}
e^{-\left( \gamma -g\right) \left\vert i-i_{0}\right\vert } & i>i_{0} \\
e^{-\left( \gamma +g\right) \left\vert i-i_{0}\right\vert } & i<i_{0}%
\end{array}%
\right. ,  \label{wavelocal}
\end{equation}%
which has different decaying behaviors on different sides of the localization
center.
When $\left\vert g\right\vert \geq \gamma $, delocalization occurs
on one side and then skin state emerges to the boundary. The
transition point from the localized state to skin state is given by
\begin{equation}
\gamma(E) =\left\vert g \right\vert . \label{LDP}
\end{equation}
Since a localized state is not sensitive to the boundary condition of the
system, we conclude that the boundary of localization-delocalization transition under the PBC is also given by Eq.(\ref{LDP}).

For  general Hermitian quasi-periodic model (\ref{Ham1}),
LEs  might depend on $E$, and mobility edge will occur. Consequently, if $g\neq0$,
Eq. \eqref{LDP} gives the mobility edge from the localized state to extended (skin) state.
The localized eigenstate in the case $g=0$ is still localized when $0<g<\gamma(E)$. However,
the localized eigenstate becomes extended (skin) under the PBC (OBC),
when $g>\gamma(E)$.

For the non-Hermitian AA model with $h\neq0$ and $g=0$, the LEs of the localized states are given by $\gamma= |h| + \ln |\lambda_1|$, according to Eq.(\ref{LE}). Thus if $g\neq 0$, by using Eq.(\ref{LDP}), the localization transition boundary is determined by
\begin{equation}
\left\vert h\right\vert +\ln \left\vert \lambda_1 \right\vert =\left\vert
g\right\vert ,  \label{LDP1}
\end{equation}
which can
be alternatively represented as
\begin{equation}
|\lambda_1|=e^{-|h|+|g|}.  \label{tran1}
\end{equation}%
While all eigenstates are localized for $|\lambda_1| >e^{-|h|+|g|}$, they become extended (skin) states under PBC (OBC)
for $|\lambda_1| < e^{-|h|+|g|}$. The model
reduces to the classical AA model when $h=0$ and $g=0$. Eq.(\ref{tran1}) recovers the
result of Ref.\cite{longhiPRL} for $h\neq 0$ and $g=0$ and the
result of Ref.\cite{jiang2019interplay} for $g\neq 0$ and $h=0$.

\begin{figure}[tbp]
\includegraphics[width=0.48\textwidth]{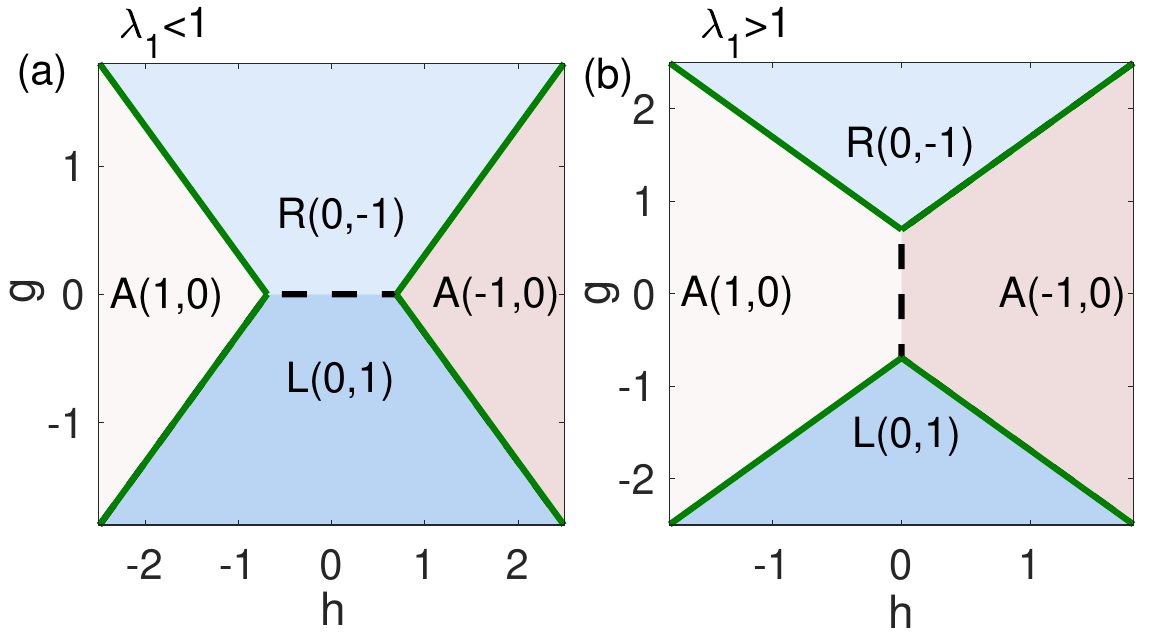}
\caption{Phase diagrams for the case with (a) $|\lambda_1|<1$ and (b) $|\lambda_1|>1$, respectively.
The phase boundaries are denoted by the green solid lines,
which are determined by $|g|=\ln |\lambda_1| + |h|$.
The winding numbers ($\nu_\phi, \nu_\psi $) are defined in the text.
$\{L,R,A\}$ represent the left-skin, right-skin, and Anderson localized phases, respectively.
The  left-skin and right-skin phases under OBC correspond to extended phases under PBC.
We have taken $\lambda_1=0.5$ in (a) and $\lambda_1=2$ in (b). }
\label{fig4}
\end{figure}

By using either Eq.(\ref{LDP1}) or Eq.(\ref{tran1}), we can obtain the complete phase diagram.
For a given $\lambda_1$, we display the phase diagram in Fig.\ref{fig4} with the phase boundaries (solid lines) determined by
Eq.(\ref{LDP1}). Fig. \ref{fig4}(a) and \ref{fig4}(b) correspond to the case  $\lambda_1<1$ and $\lambda_1>1$, respectively.
Here regions labeled by $A$ denote the Anderson localized phase, and regions labeled by $L$ or $R$ represent the
left or right skin states under OBC, which are  extended phase under PBC. From the phase diagrams, we can see that,
while increasing $|h|$ tends to driving the system into localized phase,  increasing $|g|$ tends to driving the
system into extended (skin) phase. When $|h|= |g|$, the transition point is given by $\lambda_c =1$,
which is irrelevant to the values of $h$ and $g$, and eigenstates for system with  $\lambda_1 < 1$ ($\lambda_1 > 1$)
are extended (localized).

\begin{figure}[tbp]
\includegraphics[width=0.48\textwidth]{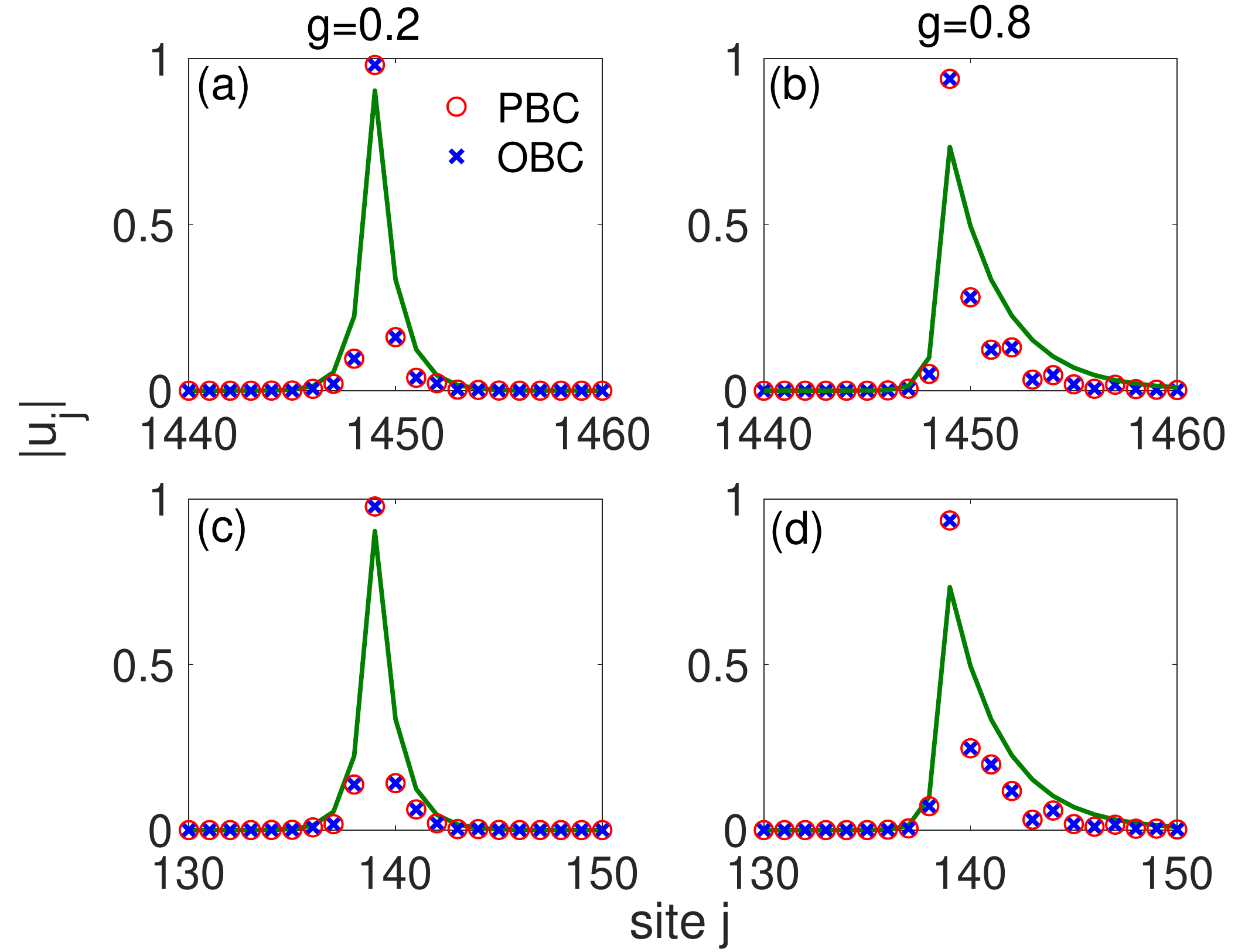}
\caption{The distribution of eigenstates for systems with $\lambda_1=2$, $h=0.5$, $N=1597$, $g=0.2$ and $0.8$, respectively.
The red circles and blue crosses represent the eigenvalues under PBC and OBC, respectively. The solid line
is plotted by using Eq. (\ref{wavelocal}).  The localization centers for states shown in (a) and (b) are at $j=1449$, and  for (c) and (d) are at $j=139$.  }
\label{fig5}
\end{figure}

For the AA model, the LE of the localized state is independent of $E$.
This suggests that all eigenstates are either localized or extended (skin) states with the
transition point independent of $E$. Also, we can conclude that all localized states of
non-Hermitian AA model can be described by a unified wavefuntion (\ref{wavelocal}) with
different states having different localization centers, as shown in Figs. \ref{fig5}(a) and (c)
or Figs. \ref{fig5}(b) and (d).

\begin{figure}[tbp]
\includegraphics[width=0.5\textwidth]{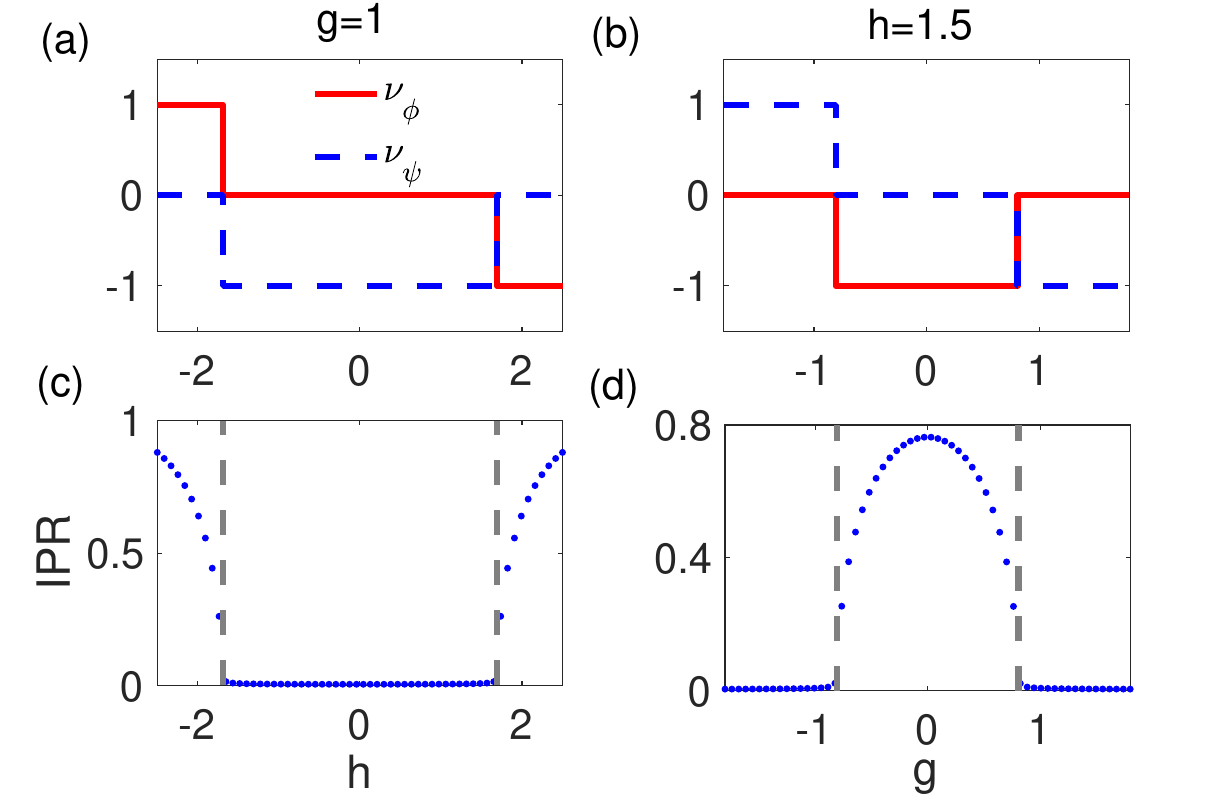}
\caption{The winding number $\nu_{\phi}$, $\nu_{\psi}$ versus $h$  for the system with  $g=1$ (a) and
versus $g$  for the system with  $h=1.5$ (b).
The IPR of eigenstates corresponding to the middlemost
real part of eigenvalue Re$(E)$ versus $h$ for the system with  $g=1$ (c)  and
versus $g$ for the system with  $h=1.5$ (d). Other parameters are $\lambda_1 =0.5$ and $N=233$,
and we have taken PBC in the numerical calculations.}
\label{fig6}
\end{figure}

\section{Winding numbers}

The phase factor $\phi$ of the potential provides a parameter space to define topological invariant, i.e. the winding number, to characterize the topological phase of the non-Hermitian quasiperiodic system. Changing the complex phase $h$ may induce topological phase transition. Recall that
the winding number of the system can be defined as
\begin{equation}
\nu _{\phi}= \frac{1}{2\pi i}\frac{1}{N} \int_{0}^{2\pi}\text{d}\phi \partial _{\phi }\ln
\det [H(\phi )-E_{B}], \label{nu-phi}
\end{equation}%
which measures the change of spectrum with respect to the base energy $E_{B}$
when $\phi $ is changed continuously.
Thus it is interesting to know where does the topological transition occur? Indeed, this question can also be answered with the help of LE.
As we explained above, the slope of LE, which is called acceleration \cite{Avila2015} is quantized.
We will  show that {\it quantization of the winding number just means the acceleration is quantized.}

Let us explain in more details.  In the  case $g=0$, as we show in the Appendix B,  based on Cauchy-Riemann equation, we have the following relation
\begin{equation} \label{windingpsi}
\nu _{\phi}(E,h)= -\frac{\partial \gamma (E, h)}{\partial h} .
\end{equation}
Here $\frac{\partial \gamma (E, h)}{\partial h}$ is exactly the  ``acceleration" as defined by  Avila \cite{Avila2015}. Note $\gamma \left( E,h\right)$ is a piece-wise linear function with respect to $h$ with their slopes being integers, then acceleration is  just the slope. In the following sections, we only need to consider $h>0$. Due to the symmetry \eqref{sym}, $\nu _{\phi} (E,h)$ always satisfies
\begin{equation}
\nu _{\phi} (E,-h)=-\nu _{\phi} (E,h).
\end{equation}

If $g\neq 0$,  then  boundary conditions make things different. Under PBC,  we will show that
 \begin{equation}\label{windingg}
\nu_{\phi} (E,h,g) = \left\{
\begin{array}{cc}
0, &~ |g|> \gamma (E, h)  \\
 -\frac{\partial \gamma (E, h)}{\partial h}, & ~ 0<|g|<\gamma (E, h)
\end{array}%
\right. .
\end{equation}
However, within OBC, we have
 \begin{equation}\label{windingg1}
 \nu_{\phi} (E,h,g) =\nu _{\phi}(E,h)= -\frac{\partial \gamma (E, h)}{\partial h},
 \end{equation}
for any $g\neq 0$.
The Eqs. (\ref{windingg}) and (\ref{windingg1}) are strictly educed in Appendix B.
The different boundary conditions correspond to the different winding number,
which can also be interpreted as a phenomenon for breakdown of bulk-boundary correspondence.

Based on Eqs.\eqref{LEG}, \eqref{windingpsi} and \eqref{windingg}, our analysis really
shows the topological transition of general non-Hermitian quasiperiodic models, that is to say,
the transition point is also determined by the LE $\chi_i(E)$ of the dual model \eqref{zlong}.
For simplicity, we will just demonstrate this with  AA model and  Soukoulis-Economou model.

For  the non-Hermitian AA model, when $\lambda_1<1$,
the Eqs. \eqref{windingpsi} and \eqref{windingg} can be expressed as
 \begin{equation}
\nu _{\phi}(E,h)= \left\{
\begin{array}{cc}
0,& ~ h< -\ln |\lambda_1|, \\
-1, &~h>-\ln |\lambda_1|,
\end{array}%
\right.
\end{equation}
and
 \begin{equation}
\nu _{\phi}(E,h,g)= \left\{
\begin{array}{cc}
0,& ~ g> h +\ln |\lambda_1|, \\
-1, &~g< h +\ln |\lambda_1|,
\end{array}%
\right.
\end{equation}
based on Eq.(\ref{LE}).
In Fig.\ref{fig6}, we show how the winding numbers and IPR change with $h$ or $g$.
Fig. \ref{fig6}(a) and (c) are for the system with fixed $\lambda_1=0.5$ and $g=1$.
According to Eq.(\ref{LDP}), we have $|h_c| = |g|-\ln |\lambda_1|\approx 1.7$.
It is shown that the winding number  $\nu_{\phi}$ takes different integer $0$ or $\pm 1$ in the region $|h|< 1.7$ or
$|h|>1.7$, and IPR shows that the corresponding states are extended or localized.
Fig.\ref{fig6}(b) and (d) show the winding numbers $\nu_{\phi}$ and IPR of the system
with $\lambda_1=0.5$ and $h=1.5$ versus $g$. According to Eq.(\ref{LDP}), we have $|g_c| = |h|+\ln |\lambda_1|\approx 0.8$.
The winding number $\nu_{\phi}$  takes different integer $0$ or $-1$ in the region $|g|< 0.8$ or
$|g|>0.8$, and IPR shows that the corresponding states are localized or extended.
The numerical results clearly indicate that the winding number changes its value when crossing the boundary
of localization transition and takes different integer in extended and localized regions. Consequently, we also
show that different phases in the phase diagram of Fig.\ref{fig4} can be characterized by different $\nu_{\phi}$.

\begin{figure}[tbp]
\includegraphics[width=0.48\textwidth]{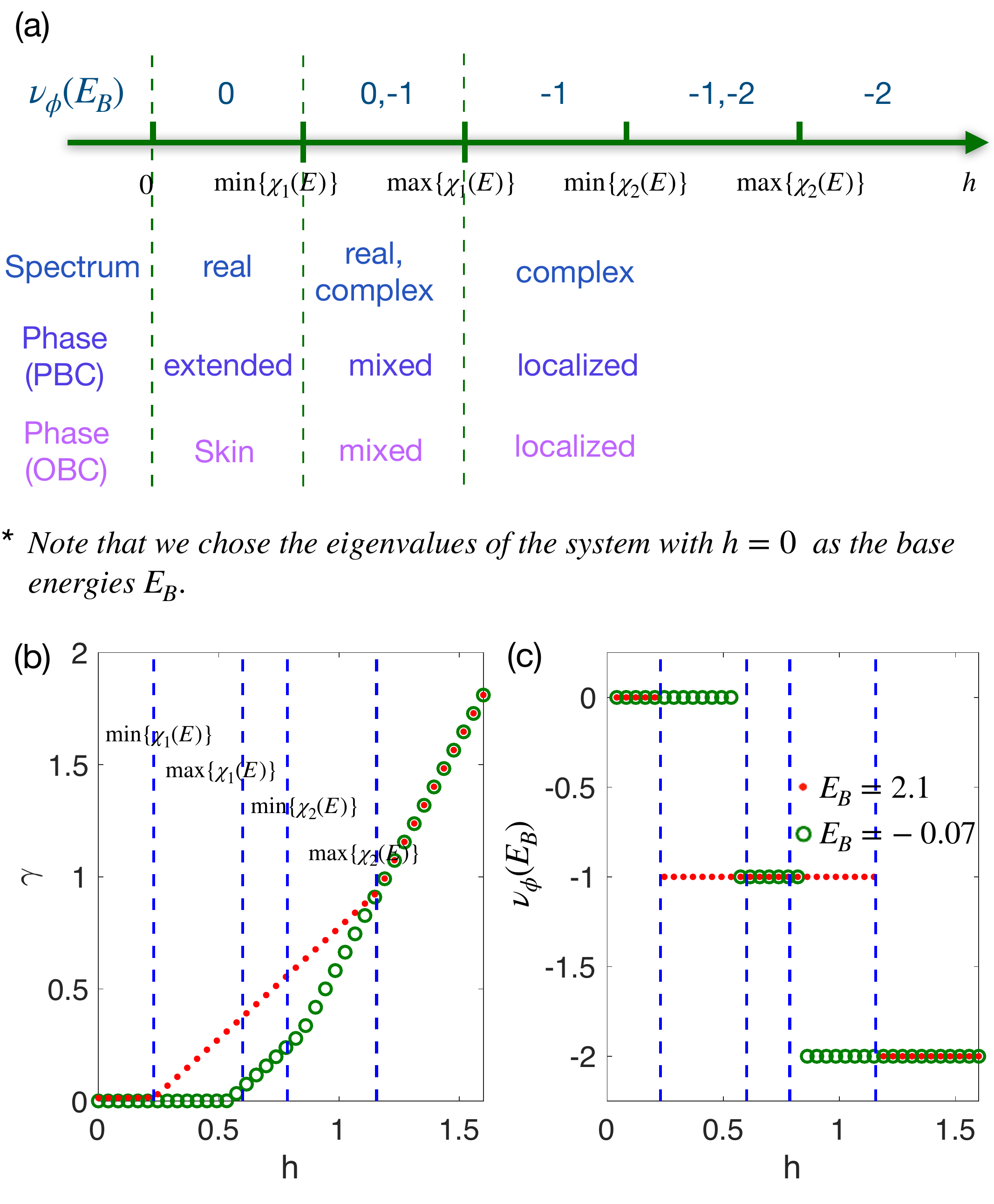}
\caption{ (a)The winding numbers $\nu_{\phi}$, spectrum and phases for the system
with potential (\ref{cos2}) and $g=0$ v.s. $h$, which is associated with the minimum and
maximum values of $\chi_{1,2}(E)$.
Numerical results for the LE (b) and the winding numbers $\nu_{\phi}$ (c)
of systems with $\lambda_1=0.2$, $\lambda_2=0.25$, $g=0$ and
$E_B=-0.071$ and $2.1$ v.s. $h$.
For this case, the minimum and maximum values of $\chi_{1,2}(E)$ are
$\min\{\chi_1(E)\}=0.23$, $\max\{\chi_1(E)\}=0.6$, $\min\{\chi_2(E)\}=0.786$, and
$\max\{\chi_2(E)\}=1.156$, respectively.}
\label{fig7}
\end{figure}

Next we turn to the Soukoulis-Economou model (\ref{cos2}).
Take the eigenvalue $E_B$ of Hermitian system as the base energy,  substitute
Eq. (\ref{LEcos2}) into Eq. (\ref{windingpsi}), then one obtains its winding number when changing the parameter $h$:
 \begin{equation}
\nu _{\phi} (E_B,h)= \left\{
\begin{array}{cc}
0,& ~ 0<h<\chi_1(E_B) ,\\
-1, &~ \chi_1(E)<h< \chi_2(E_B) ,\\
-2, &~ h>\chi_2(E_B). \\
\end{array}%
\right.
\end{equation}%
In Fig. \ref{fig7} (a), we schematically display the winding number versus the change of $h$.
Clearly, the value of winding number depends on the choice of base energy.
Fig. \ref{fig7} (b) and (c) show the numerical results of the LE and winding number as a function of $h$
with $g=0$, $\lambda_1=0.2$, $\lambda_2=0.25$, $E_B=2.1$ and $-0.07$, respectively.
When $h=0$, all eigenstates of the system are extended, thus $\gamma (E_B,0)=0$.
It is clear that each of the LE $\gamma (E_B,h)$  is a continuous piecewise linear function with variable $h$ and has two
extreme points $h=\chi_1(E_B)$ and $h=\chi_2(E_B)$, where $\chi_{1,2}(E_B)$ are the LEs of the dual model  \eqref{Dualcos2}, as shown in Fig \ref{fig3} (a).
It is obvious that the slope of the LE is zero in the region $0<h<\chi_1(E_B)$
and the winding number $\nu_{\phi}(E_B,h)=0$, all the eigenstates are extended and all the eigenenergies are real in this region.
The slope of the LE $\gamma (E_B,h)$ is $1$ and the winding number $\nu_{\phi}(E_B,h)=-1$ in the region $\chi_1(E)<h< \chi_2(E)$.
The slope of the LE  $\gamma (E_B,h)$ is $2$ and the winding number $\nu_{\phi}(E_B,h)=-2$ if $h> \chi_2(E)$. Consequently, according to $\min \{\chi_{1,2}(E_B)\}$ and $\max \{\chi_{1,2}(E_B)\}$, the parameter space
$h>0$  can be divided into five regions, as shown in Fig.\ref{fig7} (a).
In the regions $0<h<\min \{\chi_1(E_B)\}$, $\max \{\chi_1(E_B)\}<h<\min \{\chi_2(E_B)\}$, and
$\max \{\chi_2(E_B)\}<h$, the winding numbers $\nu_{\phi}(E_B,h)$ are $0$, $1$, and $2$, respectively and don't change with $E_B$.
In the region $\min \{\chi_1(E_B)\}<h<\max \{\chi_1(E_B)\}$, $\nu_{\phi}(E_B)=-1$ or $0$, which depends on
the selection of $E_B$, as shown in Fig.\ref{fig8} (c).  Indeed, not only the winding numbers, but also the states are mixed in this region.
In the region $\min \{\chi_2(E_B)\}<h<\max \{\chi_2(E_B)\}$, $\nu_{\phi}(E_B,h)=-2$ or $-1$,
which also depends on the the selection of $E_B$. Although the winding numbers are mixed, the states are not mixed,
all the eigenstates are localized in this region.

\section{ Robust spectrum and skin effect}

In Section  II, we unveiled  that for general non-Hermitian quasiperiodic models \eqref{Ham1},
if all the eigenstates of the Hermitian case ($h=0$) are extended, then
the real spectrum keeps invariant in the whole extended region $h<\min\{\chi_1(E)\}$,
i.e. the existence of robust spectrum.  In this section, we will show that this kind of
intriguing phenomenon could also happen for  the non-reciprocal hopping. We will demonstrate this with AA model.

\begin{figure}[tbp]
\includegraphics[width=0.48\textwidth]{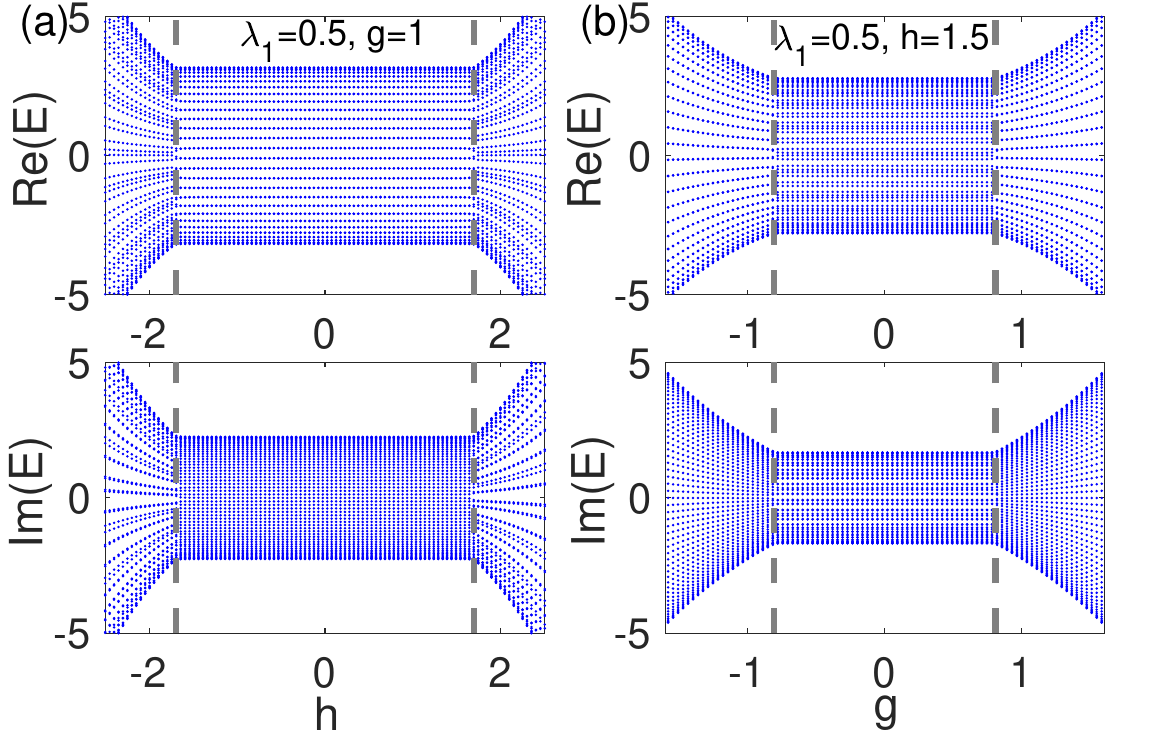}
\caption{(a) The real  and  the imaginary part of the eigenvalue spectra versus
$h$ for the AA model with $\lambda_1 =0.5$, $g=1$ and $N=55$.
(b) The real and imaginary part of the eigenvalue spectra versus
$g$ for the system with $\lambda_1 =0.5$, $h=1.5$ and $N=55$.
Dashed gray lines represent transition points.}
\label{fig8}
\end{figure}

For the AA model, and the $\mathcal{PT}$-symmetrical case with $g=0$,
the robust spectrum takes place in the regime $|h|<|h_c|= -\ln |\lambda_1|$.
On the other hand, for the  case  $h=0$, we find that robust spectrum also occurs in
the whole localized region $|g|<|g_c|=\ln |\lambda_1|$. The spectrum properties in these two limits
can be understood from the observation that the two limit cases can be related together by a dual
transformation \cite{jiang2019interplay}. For the general case with $g \neq 0$ and $h \neq 0$, with the PBC, the
spectrum is complex. Nevertheless, we find that the complex spectrum still keeps invariant when we change $h$ in
the extended region $|h|<|h_c|=|g|-\ln |\lambda_1|$ for a fixed $g$ or change $g$ in the localized region $|g|<|g_c|=|h|+\ln |\lambda_1|$ for a fixed $h$.
To give some concrete examples, we display the spectrum for the system with $\lambda_1 =0.5$ and $g=1$
versus $h$ in  Fig.\ref{fig8}(a) and the system with $\lambda_1 =0.5$ and $h=1.5$ versus $g$ in  Fig.\ref{fig8}(b).
For the case of Fig.\ref{fig8}(a), all eigenstates in the region of $|h|<1.7$ are extended state, and the corresponding
spectrum does not change with $h$ as long as $|h|<1.7$. For Fig.\ref{fig8}(b), all eigenstates in the region
of $|g|<0.8$ are localized state, and the corresponding spectrum does not change with $g$ as long as $|g|<0.8$.

Next we shall give a straightforward explanation of robust spectrum
shown in Fig.\ref{fig8}(b). In the region $|g|<0.8$, the states are
localized and are not sensitive to the boundary condition.
Therefore, the spectra under PBC and OBC should be the same in the large size limit as long as $|g|<|g_c|$.
From Eq.(\ref{smt}), we know that the open boundary spectrum is irrelevant with $g$ and should be identical
to the case of $g=0$ due to the similar transformation does not change the spectrum.  Therefore, it is not hard
to understand why the periodic boundary eigenenergies do not change with $g$ for the localized states.
When $|g|>|g_c|$, the spectra are sensitive to the boundary condition and the corresponding states are
extended or skin states under PBC or OBC.

It is not so straightforward to understand the invariance of spectrum  shown in Fig.\ref{fig8}(a).
Nevertheless, we can give an explanation by resorting the dual transformation. From this aspect, it is also useful to  consider the ring chain with a flux
penetrating through the center, yielding
\begin{eqnarray}
H(\psi ) &=&\sum_{j}(t_{L}e^{i\psi }|j\rangle \langle
j+1|+t_{R}e^{-i\psi}|j+1\rangle \langle j|  \notag \\
&&\left. +\lambda_1 \cos(2\pi \omega j+\theta)|j\rangle \langle j|\right) \label{AAp},
\end{eqnarray}%
or equivalently by replacing the hopping term connecting the first and $N$-th site as $h_{IN}=t_{L}e^{-iN \psi }|N\rangle \left\langle 1\right\vert
+t_{R}e^{iN\psi }\left\vert 1\right\rangle \langle N|$, and the winding
number is defined as
\begin{equation}
\nu _{\psi }=\frac{1}{2\pi i} \frac{1}{N} \int_{0}^{2\pi }\text{d}\psi \partial _{\psi
}\ln \det [H(\psi )-E_{B}]. \label{nu-psi}
\end{equation}
$\nu _{\psi }$ have been utilized to characterize the loop of the energy
spectra of extended and localized states \cite{longhiPRL,jiang2019interplay,Zeng2020,Gong}.

By utilizing the dual transformation:
\begin{equation*}
\left\vert j\right\rangle =\frac{1}{\sqrt{N}}\sum_{k}e^{-i2\pi \omega
kj}\left\vert k\right\rangle,
\end{equation*}%
we can get a duality form of the Hamiltonian (\ref{Ham1}) with $\lambda_{l\geqslant2}=0$, given by
\begin{equation}
\tilde{H}=\sum_{k}\left( \lambda _{L}|k\rangle \left\langle k+1\right\vert
+\lambda _{R}\left\vert k+1\right\rangle \langle k|+ t_{k}|k \rangle \langle
k |\right) ,  \label{Ham2}
\end{equation}%
where $\lambda _{L}=\lambda_1 e^{-h}$, $\lambda _{R}=\lambda_1 e^{h}$ and $%
t_{k}=2 \cos \left( 2\pi \omega k+ig\right) $. The Hamiltonian (\ref{Ham1}) with $\lambda_{l\geqslant2}=0$
and (\ref{Ham2}) have similar formulae only with coefficients difference,
but have the same spectrum, although the wave functions of two Hamiltonian are
entirely different.
Let $\lambda_1$ as the unit of the energy, we can relabel $g'=h$, $h'=g$, $\lambda'=1/\lambda_1$.
Now we can see that the case of Fig.\ref{fig8}(a) with a fixed $g$ and different $h$ can be mapped to the case with a fixed $h'$ and different $g'$, i.e., the case of Fig.\ref{fig8}(b) in the dual Hamiltonian (\ref{Ham2}). So we can apply similar explanation why the spectrum is invariant in the region of $g' < |g'_c|$ ($h<|h_c|$) for fixed $h'$ ($g$).
We note under the dual transformation, the flux phase factor $\psi$ is transformed to the phase factor $\phi'$ , i.e., $H(\psi, \lambda_1, h,g)$ is mapping to $\tilde{H}(\phi', \lambda',h',g')$.
Therefore, from the definitions of Eq.(\ref{nu-phi}) and  Eq.(\ref{nu-psi}),
we find that $\nu _{\phi,\psi }$ can be related by the following relation
\begin{equation}
\nu_{\psi} (\lambda_1, h, g) = \nu_{\phi} (1/\lambda_1, g, h), \label{dual-winding}
\end{equation}
i.e., $\nu_{\psi}$ for the system with parameter $\lambda_1$, $h$ and $g$ can be obtained from  $\nu_{\phi}$ of the corresponding system with $\lambda'=1/\lambda_1$, $h'=g$ and $g'=h$.

From the phase diagrams in Figs.\ref{fig4}, we  always have $\nu_{\phi}=0$ in the extended region
and $\nu_{\psi}=\pm 1$. Nonzero winding number $\nu_{\psi}$ indicates the existence of skin states
for the system under OBC \cite{Yokomizo,Okuma,KZhang}. On the other hand, we have always
$\nu_{\psi}=0$  in the localized region and $\nu_{\phi}=\pm 1$. The relation Eq.(\ref{dual-winding})
constructs a mapping between the phase diagram of $\lambda_1<1$ and that of $\lambda_1>1$.
The winding number $\nu_{\phi}$ ($\nu_{\psi}$) in Fig.\ref{fig4}(a) can be read out from
$\nu_{\psi}$ ($\nu_{\phi}$)  in Fig.\ref{fig4}(b) and vice versa.

By comparing the dual Hamiltonian (\ref{Ham2}) with the original Hamiltonian (\ref{Ham1}) with $\lambda_{l\geqslant2}=0$,
we can see the existence of a self-duality point at $g=h$ and $\lambda_1=1$. At this self-duality
point, $\lambda_c=1$ is usually taken as the localization-delocalization transition point \cite{Zeng2020}.
From Eq.(\ref{tran1}), we have seen that $\lambda_c =1$ is a transition point when $|h|=|g|$,
i.e., the self-duality relation is only a special case of our general result Eq.(\ref{tran1}). It is worth
indicating that our analytical result Eq.(\ref{tran1}) does not rely on the self-duality relation or even
the dual transformation.

\begin{figure}[tbp]
\includegraphics[width=0.48\textwidth]{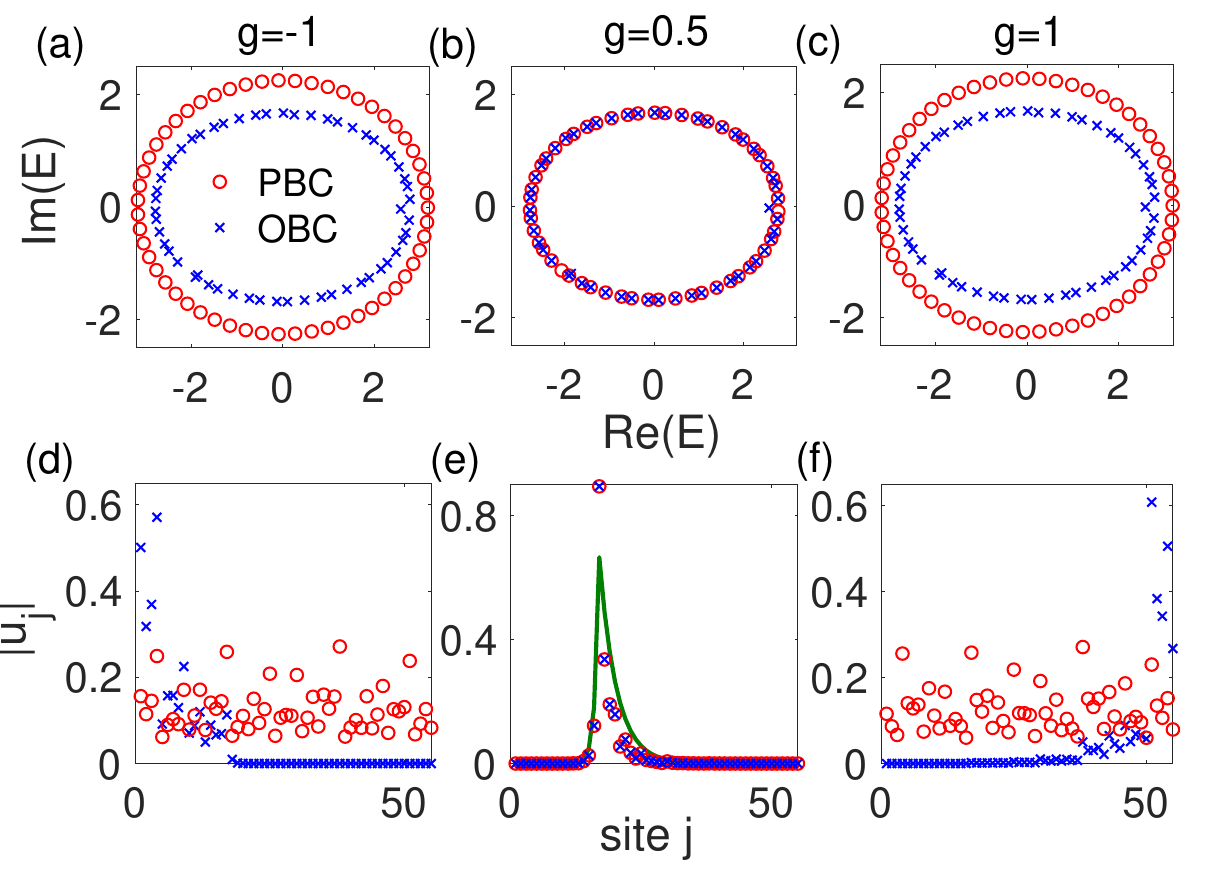}
\caption{ (a)-(c) The complex spectrum for systems with $\lambda_1=0.5$, $h=1.5$, $N=55$ and
$g=-1$, $0.5$ and $1$, respectively. The red circles and blue crosses represent the eigenvalues under PBC
and OBC, respectively. (d)-(f) The distribution of eigenstates corresponding to the minimum real part of eigenvalues
for systems with $\lambda_1=0.5$, $h=1.5$, $g=-1$, $0.5$ and $1$, respectively. The solid line in (e) is plotted by using Eq.(\ref{wavelocal}).
}
\label{fig9}
\end{figure}

Next we compare the spectra of the system under PBC and OBC to see the sensitivity of
spectra to the change of boundary conditions. If non-Hermitian skin effect exists, the system
shall display remarkably different eigenspectra under PBC and OBC \cite{Okuma,KZhang,Lee,Yokomizo,Slager}.
In the Fig.\ref{fig9}(a)-(c), we show the spectra in the complex space spanned by Re$(E)$ and Im$(E)$ for
systems with $\lambda_1=0.5$, $h=1.5$ and
$g=-1$, $0.5$ and $1$, respectively, under both PBC and OBC. As shown in Fig.\ref{fig9}(b), in the localized
region the spectrum under PBC and OBC are almost the same except for several isolated points corresponding
to edge states. On the other hand, the spectra under PBC and OBC are obviously different in the delocalized region
as  shown in Fig.\ref{fig9}(a) and \ref{fig9}(c), which is a signature for the existence of skin effect under OBC
as witnessed in the distributions of eigenstates shown in  Fig.\ref{fig9}(d) and \ref{fig9}(f), respectively. The
distributions of localized states under PBC and OBC are identical as shown in Fig.\ref{fig9}(e), showing clearly
that the localized states are independent of the boundary conditions.
The numerical results also indicate that the distributions of localized states can be well described by Eq.(\ref{wavelocal}).

\section{Application to other models}

\subsection{Generalized Ganeshan-Pixley-Das Sarma model}
\begin{figure}[tbp]
\includegraphics[width=0.48\textwidth]{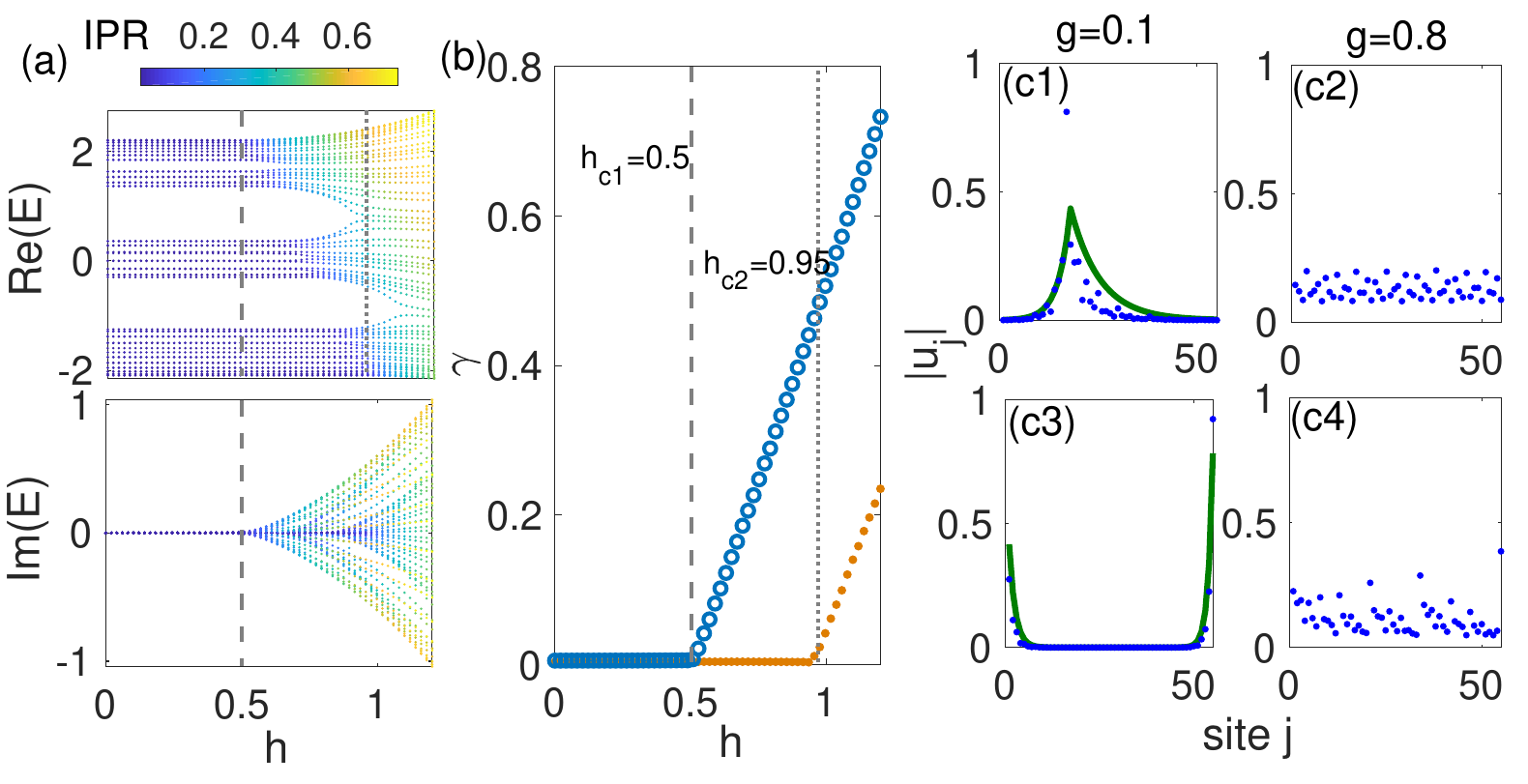}
\caption{ (a) The real and the imaginary part of the eigenvalue spectra versus $h$ for the system with potential (\ref{Vfrac}).
Here $g=0$, $\lambda=0.5$, $b=0.1$ and $N=55$.
 (b)The LE of eigenstates corresponding to the minimum
 (circles) and maximum(dots) real part of eigenvalues. For accuracy of LE, we choose $N=1597$.
  (c1) and (c2) The distribution of eigenstates corresponding to the minimum real part of eigenvalues
  for systems with $\lambda=0.5$, $h=1.2$, $g=0.1$ and $0.8$,
 respectively. (c3)(c4) The distribution of eigenstates corresponding to the maxmum real part of eigenvalues for systems with
 $\lambda=0.5$, $h=1.2$, $b=0.1$, $g=0.1$ and $0.8$, respectively.  The solid lines
 in (c1) and (c3) are plotted by using Eq.(\ref{wavelocal}). In (c), $N=55$.}
\label{fig10}
\end{figure}

Next we consider the  generalized complex Ganeshan-Pixley-Das Sarma model \cite{ganeshan2015nearest}
 \begin{equation}
V_j=2\lambda \frac{ \cos (2\pi \omega j+i h)}{1-b\cos (2\pi \omega j+i h) }, \label{Vfrac}
\end{equation}
which is the first quasiperiodic model that the mobility edges have analytic formula for the Hermitian case$(g,h=0)$.
By applying Avila's global theory, the LE of the non-Hermitian model can be easily derived, and the expression is
\begin{equation}\label{inequation1.11}
\gamma(E,h)\ge\max\{|h|+\ln \frac{ |b  E+2\lambda|+\sqrt{( b  E+2\lambda)^2-4 b ^2}}{2(1+\sqrt{1-b^2})},0\},
\end{equation}
when $|h|<\ln|\frac{1+\sqrt{1- b ^2}}{ b }|$.
The slope of $\gamma(E,h)$ might be $\pm 1$ or $0$.
Fig.\ref{fig10} (a) shows the spectrum for the system with $g=0$, $\lambda=0.5$ and $b=0.1$
versus $h$. The spectrum does not change with $h$ in the extended-state region: $h<0.5$. This indicates clearly the existence of robust spectrum in the complex Ganeshan-Pixley-Das Sarma model.
For $h>0.5$, the eigenstates with high real part of energies become localized firstly, thus the LE
of eigenstates corresponding to the minimum and maximum real part of eigenvalues
can determine the mobility edge region, as shown in Fig.\ref{fig10} (b). In the region $h>0.95$,
all eigenstates are localized. Then we consider the system with $g \neq 0$. The wavefunction (\ref{wavelocal})
and the transition point (\ref{LDP}) tell us that $|g|<\gamma$, the states keep localized, while
$|g|>\gamma$, the states become extended, where $\gamma$ is the LE with $g=0$. Fig.\ref{fig10} (c) shows the distribution of
eigenstates corresponding to the minimum and maximum real part of eigenvalues for systems
with $h=1.2$. Eigenstates are localized with $g=0.1$ and are extended with $g=0.8$, as shown in Fig.\ref{fig10} (c).

\subsection{Quasiperiodic exponential potential}
The non-reciprocal hopping model with the quasiperiodic exponential potential
\begin{equation}
V_{j}= V e^{ i ( 2 \pi \omega j+\phi)},
\end{equation}
has the same basic idea to determine the transition point (\ref{LDP}).
The LE of the localized states for this model with $g=0$ is $\gamma=\ln(V)$,
so the boundary of localization transition is given by
\begin{equation}
|V|= e^{ \left\vert g\right\vert}. \label{Vg}
\end{equation}
The full details for the calculation of the LE  are given in Appendix.
While all eigenstates are localized for $|V|> e^{ \left\vert g\right\vert}$, the eigenstates are
extended states (skin states) under PBC (OBC) for $|V|< e^{ \left\vert g\right\vert}$. When
$g=0$, the model reduces to the one studied in Ref.{\cite{longhi2019metal}} and no skin
effect occurs. For $g \neq 0$, skin effect occurs in the region of $|V|< e^{ \left\vert g\right\vert}$.

The unusual spectrum feature can be also found in this model. Eq.(\ref{Vg}) suggests that the localized
phase exists only for $|V|>1$. For a given $V$ with $|V|>1$, the system is in localized phase in the region
$|g|<g_c$ with $g_c=\ln |V|$. We find that the spectrum of the system is invariant with the change of $g$
as long as $|g|<g_c$, which is verified by our numerical result and can be explained in a similar way as
given in the above subsection.

\section{Summary and outlook}

In summary, based on Avila's global theory, one of his Fields medal work,
we developed a rigorous and general scheme for the study
of non-Hermitian quasiperiodic systems with both complex phase factor and
non-reciprocal hopping.  We demonstrated that the localization-delocalization transition point, $\mathcal{PT}$-symmetry-breaking point for general non-Hermitian quasicrystals with $h \neq 0$, can be described by a conclusion expression $h=\min\{\chi_1(E)\}$, where $\chi_1(E)$ is the
smallest positive LE of its dual model with $h=0$ for a given eigenvalue $E$.
The general relation between winding numbers and acceleration is also unveiled, consequently we obtained that
the winding number is just the slope of LE,
and the topological transition points for the winding numbers are determined by
all dual-model LEs $h=\chi_i(E)$. These results are applied to study the typical examples, including both the non-Hermitian AA model and Soukoulis-Economou model. Especially, for the  non-Hermitian AA model, we analytically determined the complete phase diagram in the whole parameter space, which can be
alternatively characterized by winding numbers. Moreover, we discovered an intriguing feature of robust spectrum, i.e., the spectrum
keeps invariant under the change of the complex phase parameter
$h$ or non-reciprocal parameter $g$ as long as $h<|h_c|$ or $g<|g_c|$
for system in the extended or localized region, respectively. We found that the
existence of robust spectrum is a very common feature of non-Hermitian quasiperiodic systems.
Models beyond the two typical examples are also discussed.
Our analysis open a door to further study  intriguing properties of non-Hermitian quasicrystals.

Photonic systems provide a valid platform of realization of
non-Hermitian Hamiltonians with quasiperiodic potentials, which are manifested by the gain and loss of the
laser pulse inside the optic fiber. Many typical phenomena have been observed in photonic experiments,
such as PT symmetry, exceptional points, non-Hermitian skin effect., etc.
\cite{Regensburger11,Regensburger12,Wimmer15,Vatnik17,Derevyanko19,Weidemann20}.
We expect that our theoretical work shall stimulate the experimental study of localization transition in non-Hermitian quasiperiodic systems.

\begin{acknowledgments}
This work is supported by the National Key
Research and Development Program of China (2016YFA0300600), NSFC under Grants Nos. 11974413 and the Strategic
Priority Research Program of CAS (XDB33000000). Q. Zhou was  supported by
 National Key R\&D Program of China (2020YFA0713300), NSFC grant (12071232), The Science Fund for Distinguished Young Scholars of Tianjin (No. 19JCJQJC61300)\end{acknowledgments}

\appendix

\section{Global theory of one-frequency cocycle }
Suppose that $A$ is an analytic  function from the circle $S^{1}$ to the group $SL(2,C)$, an
analytic  quasiperiodic cocycle $(\omega,A)$ can be seen as a linear skew product:
\begin{equation*}
(\omega,A):  S^{1}\times R^{2} \rightarrow S^{1}\times R^{2}
\end{equation*}
\begin{equation*}
 (\theta,v)\mapsto(\theta+\omega,A(\theta)\cdot v).
\end{equation*}
If $A(\theta)$ admits a holomorphic extension to $| Im \theta|<\delta$, then for $|\epsilon|<\delta$
we can define $A_{\epsilon}(\theta)=A(\theta+i\epsilon)$, and define its LE by
\begin{equation}
\gamma \left( E,h\right) =\lim_{n\rightarrow \infty }\frac{1}{ 2 \pi n}\int_{0}^{2 \pi} \ln
\left\vert \left\vert T_{n}\left( E,\phi ,h\right) \right\vert \right\vert
d\phi ,
\end{equation}
where $T_{n}$ is the transfer matrix. The key observation of Avila's global theory is that
$h\rightarrow \gamma \left( E,h\right)$ is convex and piecewise linear, with right-derivatives
satisfying
\begin{equation*}
\lim_{h \rightarrow 0_{+} }\frac{1}{2\pi h }(\gamma \left( E,h\right)-\gamma \left( E,0\right))\in \mathbb{Z}.
\end{equation*}
Similarly, the left-derivative satisfy
\begin{equation*}
\lim_{h \rightarrow 0_{-} }\frac{1}{2\pi h }(\gamma \left( E,h\right)-\gamma \left( E,0\right))\in \mathbb{Z}.
\end{equation*}

Note a sequence $(u_n)_{n \in \mathbb{Z}}$ is a formal solution of the
eigenvalue equation $$u_{n+1}+u_{n-1}+V(\theta+n\omega)u_n=Eu_n$$ if and only if
it satisfied $$\begin{pmatrix}
u_{n+1}\\u_n\end{pmatrix}= \begin{pmatrix}
 E-v(\theta+n\omega) &  -1\cr
  1 & 0 \end{pmatrix} \begin{pmatrix} u_n\\u_{n-1} \end{pmatrix},$$
  therefore, any quasiperiodic model \eqref{Ham1} can be seen as a quasi-periodic cocycle.

Generally speaking, it is difficult to exactly calculate the LE, however Avila's global theory actually provide an efficient way to calculate the LE,
thus to  determine to localize-delocalize transition. In the following, we will illustrate this by two well-known models, and explain the general results:

\subsection{AA model}

The eigenvalue equation of AA model is
$$u_{n+1}+u_{n-1}+2\lambda_1 \cos 2\pi (\theta+n\omega)u_n=Eu_n,$$
thus the corresponding cocycle is $(\omega, T(\theta))$ where
$$T(\theta)= \begin{pmatrix}
 E-2\cos \theta &  -1\cr
  1 & 0 \end{pmatrix}. $$
Let us complexify the phase, and let $h\rightarrow +\infty$,  direct computation shows that
\begin{equation*}
T(\phi+i h)=e^{ h}e^{i2\pi(\theta+ \omega)}\left(
\begin{array}{cc}
 -\lambda &  0 \\
0 & 0 \\
\end{array}
\right) + o(1).
\end{equation*} Thus we have  $
\gamma \left( E,h\right)= h +\log|\lambda| +o(1).$
Note
 $\gamma (E,h)$ is a convex, piecewise linear function of $h$ with their
slopes being integers, thus if $h$ is large enough,
$$\gamma \left( E,h\right)= h +\log|\lambda|.$$
Furthermore,  $E$ does
not lie in the spectrum of the Hamiltonian $H$, if and only if $\gamma (E,h)>0$,
and $\gamma (E,h)$ is a linear functions around $h$.
Thus if the energy $E$ lies in the spectrum, we have
$$\gamma \left( E,h\right)= \max\{\ln| \lambda |+h, 0\}, \forall h\geq 0.$$
Note $T(\theta)\in SL(2,\mathbb{R})$, thus the LE is an even function with respect to $h$, which gives
$$\gamma \left( E,h\right)= \max\{\ln| \lambda |+|h|, 0\}, \forall h\in \mathbb{R}.$$

\subsection{Complex quasiperiodic potential}
For the complex quasiperiodic model $
V_{n}=  V e^{- i ( 2 \pi \omega n+\phi)}, $ the transfer matrix takes the form
\begin{equation*}
T(\phi)=\left(
\begin{array}{cc}
 E-  V e^{ i ( 2 \pi \omega +\phi)}&  -1 \\
1 & 0 \\
\end{array}
\right),
\end{equation*}
thus $T(\phi)\in SL(2,\mathbb{C})$, not belongs to $SL(2,\mathbb{R})$ anymore, thus compared to the AA model, the LE is not  an even  function with respect to $h$.
Still if we complexify the phase, and let $h\rightarrow +\infty$,  direct computation shows that
\begin{equation*}
T(\phi+i h)=e^{ h}e^{- i2\pi(\phi+ \omega)}\left(
\begin{array}{cc}
 -V &  0 \\
0 & 0 \\
\end{array}
\right) + o(1).
\end{equation*}
Thus we have  $
\gamma \left( E,h\right)= h +\log|V| +o(1),  h\geq 0$. Note
 $\gamma (E,h)$ is a convex, piecewise linear function of $h$ with their
slopes being integers, then if the energy belongs to the spectrum, then
$$\gamma \left( E,h\right)= \max\{\ln| \lambda |+h, 0\}, \forall h\geq 0,$$
consequently, we have
$$\gamma \left( E,0\right)= \max\{\ln| V |, 0\}.$$

On the other hand, for the model $
V_{n}=  V e^{ i ( 2 \pi \omega n+\phi)}, $ we can omplexify the phase, and let $h\rightarrow -\infty$, which gives us
 $
\gamma \left( E,h\right)= -h +\log|V| +o(1),  h\leq 0$. Similarly, we have
\begin{equation}
\gamma \left( E,0\right)= \max\{\ln| V |, 0\}. \label{CLE}
\end{equation}

\subsection{General models}

As we mentioned above, one of the fundamental results of Avila's global theory  is that $\gamma \left( E,h\right)$
is a piecewise affine function in $h$ for each $E$, and the slope of each piece is an integer.
Indeed, as proved in \cite{GJYZ}, one can give the exact turning points of $\gamma \left( E,h\right)$
and the exact slope of it in every piece, which can be seen as quantitative version of Avila's global theory,  in the following, we will try to explain this:

For any trigonometric polynomials
$$V_{j}=\sum_{k=1} 2 \lambda_l \cos(2l\pi \omega j), $$
consider the quasi-periodic model
\begin{equation}\label{schg}
Eu_j=u_{j+1}+u_{j-1}+V_j u_j,\ \ j\in\mathbb{Z}.
\end{equation}
Through the transformation
$$u_j=\sum_{k}e^{i 2 \pi \omega k j} \tilde{u}_k, \ \ k\in \mathbb{Z},$$
the dual model has the form
\begin{equation}\label{long}
E\tilde{u}_k=\sum\limits_{l=-d}^{d} \lambda_{|l|}\tilde{u}_{k+l}
+2\cos(2\pi \omega k)\tilde{u}_k.
\end{equation}
The model (\ref{long}) can be written as the following form
\begin{equation}\label{long1}
\tilde{u}_{k+d}=\frac{1}{\lambda_d}\big\{ \left [E-2\cos(2\pi \omega k)\right] \tilde{u}_{k}
-\sum\limits_{l=-d}^{d-1} \lambda_{|l|} \tilde{u}_{k+l}\big\}.
\end{equation}
So the matrix form for the model (\ref{long1}) can be written as
 \begin{equation*}
\left(
\begin{array}{c}
\tilde{u}_{k+d} \\
 \vdots \\
\tilde{u}_{k+1}\\
\tilde{u}_{k}\\
 \vdots \\
\tilde{u}_{k-d+1}\\
\end{array}%
\right) =A^{k}
\left(
\begin{array}{c}
\tilde{u}_{k+d-1} \\
 \vdots \\
\tilde{u}_{k}\\
\tilde{u}_{k-1}\\
 \vdots \\
\tilde{u}_{k-d}\\
\end{array}%
\right),
\end{equation*}
where
\begin{equation*}
A^{k}=
\left(
\begin{array}{cccccc}
\frac{-\lambda_{d-1}}{\lambda_{d}}&\cdots&\frac{E-2\cos(2\pi \omega k)}{\lambda_{d}}&\frac{-\lambda_{1}}{\lambda_{d}}&\cdots &-1\\
1 & \cdots & 0 & 0  & \cdots &0  \\
 0&\ddots & & &  & 0\\
\vdots &  &1 & &   & \vdots\\
 & & & \ddots & & \\
 0& \cdots & & & 1 & 0\\
\end{array}%
\right)
\end{equation*}
is a $2d\times 2d$ matrix. Then we can define the matrix
\begin{equation}
\mathbf{\Theta}=\left( T_{N}^{\dag }
T_{N} \right) ^{1/(2N)} , \label{NLE3}
\end{equation}
where $T_N=\prod_{k}A^{k}$ is the total transfer matrix.
When $N\rightarrow \infty$, $\mathbf{\Theta}$ is  finite, which can be guaranteed by
Oseledec's ergodic theorem.  The LEs are
\begin{equation*}
\chi_i=\ln \theta_i,
\end{equation*}
where $\theta_i$ are the eigenvalues of matrix $\mathbf{\Theta}$.

It is easy to check that
\begin{equation} \label{symplectic}
T^{dn}= \prod_{k=dn}^{dn+d-1}A^{k}=
\left(
\begin{array}{cc}
C_d^{-1}(EI-B_{dn})&-C_d^{-1}C_d^*\\
I_d & O_d  \\
\end{array}%
\right),
\end{equation}
where
\begin{equation*}
C_d=
\left(
\begin{array}{ccc}
\lambda_d&\ldots&\lambda_1\\
0& \ddots&\vdots \\
0& 0 &\lambda_d\\
\end{array}%
\right),
\end{equation*}
$C_d^*$ is its adjoint and $B_{dn}$ is the Hermitian matrix
\begin{equation*}
B_{dn}=
\left(
\begin{array}{cccc}
W_{dn+n-1} & \lambda_1&\ldots&\lambda_{d-1}\\
 \lambda_1& \ddots & \ddots &\vdots \\
\vdots&  \ddots &W_{dn+1}&\lambda_1\\
\lambda_{d-1}&\ldots& \lambda_1 &W_{dn}\\
\end{array}%
\right),
\end{equation*}
where $W_k=2\cos(2\pi \omega k)$, $I_d$ and $O_d$ are the
$d$-dimensional identity and zero matrices.

Note  the matrix (\ref{symplectic}) is
complex symplectic,  thus  the eigenvalues of (\ref{NLE3}) will come in pairs, then the
LEs can be denoted by $\pm\chi_1, \cdots, \pm\chi_\ell$ with
multiplicity $n_1, \cdots, n_\ell$ respectively. We may assume that
$0\le \chi_1(E)<\cdots <\chi_\ell(E)$. It is obvious that   $n_1+ \cdots+ n_\ell =d$.

As proved in \cite{GJYZ}, the LEs for the model (\ref{Ham1}) with $g=0$ can be written as
the following:
\begin{widetext}
\begin{eqnarray}
\gamma \left( E,h\right)=
\left\{
\begin{array}{cc} \label{gammah}
\gamma(E,0), &~~h \in [0,\chi_{1}(E)],\\
   \vdots  & \vdots \\
\gamma(E,\chi_{i}(E))+ (h-\chi_{i}(E))\sum_{j=1}^{i}{n_{j}},&~~h \in (\chi_i(E),\chi_{i+1}(E)],\\
 \vdots  & \vdots\\
\gamma(E,\chi_{\ell}(E))+(h-\chi_l(E))\sum_{j=1}^{\ell}{n_{j}}, &~~h \in (\chi_\ell(E),\infty),
\end{array} %
\right.
\end{eqnarray}
\end{widetext}
where $E\in\mathbb{C}$ and $1< i< \ell$.
For the $\mathcal{PT}$-symmetrical case: the model (\ref{Ham1}) with $g=0$,
the boundaries of the extended-mixed transformation and mixed-localized
transformation can be determined by $h=\min(\chi_{1}(E))$ and $h=\max(\chi_{1}(E))$,
which only depends on $\chi_{1}(E)$ in connection with longest localization length for the
dual model (\ref{long1}).

\section{Winding number $\nu _{\phi}$ for $g=0$ and $g\neq 0$}
The definition of winding number $\nu _{\phi}$ is
\begin{eqnarray}
\nu _{\phi}(g)&=&\lim_{N \rightarrow \infty }\frac{1}{2\pi i}\int_{0}^{2\pi}\text{d}\phi \partial _{\phi }
\frac{\ln \det [H(\theta,g)-E_{B}]}{N}\notag\\
&=&\lim_{N \rightarrow \infty }\frac{1}{2\pi i}\int_{0}^{2\pi}\text{d}\phi \partial _{\phi }
\zeta(E_B,\theta,g), \label{NUP}
\end{eqnarray}
where  $\theta=\phi+ih$ and the analytical function
\begin{equation}
\zeta(E_B,\theta,g)=\frac{\ln D_n(E_B,\theta,g)}{N}  \label{BFUN}.
\end{equation}%
with
\begin{eqnarray*}
D_n(E_B,\theta,g)=\det|H(\theta,g)-E_B|.
\end{eqnarray*}
According to the Cauchy-Riemann equation in complex form,
we can get
\begin{equation}
\frac{\partial \zeta(E_B,\theta,g)}{\partial h}=i \frac{\partial \zeta(E_B,\theta,g)}{\partial \phi}.
\end{equation}%
Then we can get
\begin{eqnarray*}
\nu _{\phi}&=&\frac{1}{2\pi i} \lim_{N \rightarrow \infty } \int_{0}^{2\pi }\text{d}\phi \partial _{\phi }\zeta(E_B,\theta,g)\\
&=&-\frac{1}{2\pi }\lim_{N \rightarrow \infty } \int_{0}^{2\pi }\text{d}\phi \partial _{h }\zeta(E_B,\theta,g)\\
&=&-\frac{1}{2\pi } \lim_{N \rightarrow \infty } \partial _{h } \int_{0}^{2\pi }\text{d}\phi \zeta(E_B,\theta,g).\\
\end{eqnarray*}%
When $g=0$ and $n \rightarrow \infty$,  the normal of the transfer matrix is
$$\lim_{n \rightarrow \infty }\frac{ \ln || T_n (E_B,\theta) ||}{N}= \lim_{n \rightarrow \infty } \zeta(E_B,\theta,0).$$
The transfer matrix of the system with $g=0$ can be written as
\begin{equation*}
T_{n}\left( E,\theta \right) =\prod_{j=1}^{n}T^{j}=\prod_{j=1}^{n}\left(
\begin{array}{cc}
E-V_{j} & -1 \\
1 & 0%
\end{array}%
\right) .
\end{equation*}%
The transfer matrix can also be expressed
\begin{equation*}
T_{n}\left( E,\theta \right) =\left(
\begin{array}{cc}
D_n(E_B,\theta,0) & -D_{n-1}(E_B,\theta,0) \\
D_{n-1}(E_B,\theta,0) &-D_{n-2}(E_B,\theta,0).%
\end{array}%
\right) .
\end{equation*}%
Then we can get
\begin{eqnarray*}
&&\lim_{N \rightarrow \infty } \int_{0}^{2\pi }\text{d}\phi \zeta(E_B,\theta,0)\\
&=& \lim_{N \rightarrow \infty } \int_{0}^{2\pi }\text{d}\phi
\frac{ \ln || T_N (E_B,\theta) || }{N} \\
&=&2 \pi\gamma (E_B,h). \\
\end{eqnarray*}%
Finally we get the relaiton
\begin{eqnarray*}
\nu _{\phi}(0)=-\partial _{h } \gamma (E_B,h).
\end{eqnarray*}%

 Now, we calculate the winding number of system with $g\neq 0$.
 The Hamiltonian with a general boundary condition in matrix form can be written as
 \begin{equation} \label{Hmatrix}
H\left(\phi,g \right)  =\left(
\begin{array}{ccccc}
V_1& 1& & &\eta e^{Ng} \\
1 &V_2& 1 & & \\
 &\ddots &\ddots& \ddots&\\
 & &1 &V_{N-1}& 1\\
\eta e^{-Ng} & & & 1 &V_N\\
\end{array}%
\right).
\end{equation}%
where $\eta$ is a finite value. $\eta=0$ or $1$ is corresponding to PBC or OBC.
In the large $N$ limit, the determinant of $H\left(\phi \right)$ is
\begin{eqnarray*}
&&\det H\left(\phi, g\right)\\
&=& (-1)^{N+1}\eta e^{N|g|}-2\times (-1)^{N+1} +\det H\left(\phi, 0\right).
 \end{eqnarray*}
First we calculate the  integrand
\begin{eqnarray*}
&&\frac{1}{2\pi i}\partial _{\phi }
\frac{\ln [D(E_B,\theta,g)]}{N}=\frac{1}{2\pi i}
\frac{\partial _{\phi } D(E_B,\theta,0)}{N D (E_B,\theta,g) }\\
&=&\frac{1}{2\pi i} \frac{\partial _{\phi } D(E_B,\theta,0)}{N ( (-1)^{N+1}\eta e^{Ng}+D(E_B,\theta,0))}.
 \end{eqnarray*}
 To calculate the above equation, we need to get the behavior of $\partial _{\phi } D(E_B,\theta,0)$
 and $D(E_B,\theta,0)$.
 According to the definition of the LE,
\begin{equation}
\lim_{N\rightarrow\infty}\ln |D(E_B,\theta,0)|/N=\gamma \label{Gam}
\end{equation}%
and thus $|D(E_B,\theta,0)|$ can be written as
 \begin{equation*}
|D(E_B,\theta,0)|=e^{\gamma N}.
\end{equation*}%
According to the definition of the winding number,
  \begin{equation*}
\lim_{N \rightarrow \infty }\frac{1}{2\pi i}\int_{0}^{2\pi}\text{d}\phi \partial _{\phi }
\frac{\ln \det [H(\theta,0)-E_{B}]}{N}=\nu_{\phi}(0),
\end{equation*}%
we can get
 \begin{eqnarray*}
\frac{ \partial _{\phi } D(E_B,\theta,0)}{D(E_B,\theta,0)}= N\Omega(\psi),
\end{eqnarray*}%
with
 \begin{eqnarray*}
\lim_{N \rightarrow \infty }\int_0^{2\pi}\Omega(\psi)= 2\pi  \nu_{\phi}(0) i.
\end{eqnarray*}%
Then, we can get
\begin{eqnarray}
&&\frac{1}{2\pi i}\partial _{\phi }
\frac{\ln [D(E_B,\theta,g)]}{N}\notag \\
&=&\left\{
\begin{array}{cc}
0,& |g|>\gamma \\
\Omega(\psi),&|g|<\gamma %
\end{array}%
\right. , \label{INDE}
\end{eqnarray}
for $\eta \neq 0$ and
\begin{eqnarray} \label{INDE2}
\frac{1}{2\pi i}\partial _{\phi }
\frac{\ln [D(E_B,\theta,g)]}{N}=\Omega(\psi),
\end{eqnarray}
for $\eta=0$.
Finally, substitution of Eq. (\ref{INDE}) into Eq. (\ref{NUP}) gets the winding number of system with
$g\neq0$ and  $\eta \neq 0$
\begin{align}
\nu _{\phi}(g)= \notag
 &\lim_{N\rightarrow \infty }\frac{1}{2\pi i}\int_{0}^{2\pi}\text{d}\phi \partial _{\phi }
\frac{\ln \det [H(\theta,g)-E_{B}]}{N}\\
=&\left\{
\begin{array}{cc} \label{vpsi}
0,& |g|>\gamma \\
\nu_{\phi}(0)=-\partial _{h} \gamma (E_B,h),&|g|<\gamma .%
\end{array} %
\right.
\end{align}%
Substitution of Eq. (\ref{INDE2}) into Eq. (\ref{NUP}) get the winding number of system under OBC ($\eta=0$)
\begin{align}
\nu _{\phi}(g) = \nu_{\phi}(0).%
\end{align}%
Thus we can get that the winding number remains unchanged under the different boundary conditions except OBC.

\end{document}